\documentclass[useAMS,usenatbib]{mn2e}
\usepackage{eso-pic,graphicx}
\usepackage{float}
\usepackage{color}
\usepackage{xspace}

\usepackage{color,multirow}
\usepackage{lscape}
\usepackage{soul}

\usepackage[all]{hypcap}
\usepackage{amsmath}
\usepackage{trackchanges}

\usepackage[normalem]{ulem}

\def\arcsec{\hbox{$^{\prime\prime}$}\xspace}
\def\arcmin{\hbox{$^{\prime}$}\xspace}
\newcommand{\ms}{$M_{\odot}$}

\newcommand{\kms}{km~s$^{-1}$}

\newcommand{\hst}{{\it \hst}}
\newcommand{\ec}{$\eta$\,Car}

\title[MN-18-1804-MJ]{\textcolor{black}{
Distinguishing Circumstellar from Stellar Photometric Variability in Eta Carinae\\
}
}

\author[Damineli et al.]
  {A.~Damineli$^{1}$\thanks{E-mail: augusto.damineli@iag.usp.br},
   E.~Fern\'andez-Laj\'us$^{2,3}$, 
   L.~A.~Almeida$^{1,6}$, 
   M.~F.~Corcoran$^{7,8}$,\\
  \newauthor
   D.~S.~C.~Damineli$^{14}$,
   T.~R.~Gull$^{9}$,
   K.~Hamaguchi$^{7}$,
   D.~J.~Hillier$^{10}$,
   F.~J.~Jablonski$^{5}$,\\
  \newauthor
  T.~I.~Madura$^{11}$,
   A.~F.~J.~Moffat$^{4}$,
   F.~Navarete$^{1}$,
   N.~D.~Richardson$^{12}$,
   G.~F.~Ruiz$^{1}$,\\
  \newauthor 
   N.~E.~Salerno$^{2}$,
   M.~C.~Scalia$^{2,3}$,
   G.~Weigelt$^{13}$ \\
  $^{1}$~Instituto de Astronomia, Geof\'isica e Ci\^encias Atmosf\'ericas da USP, Rua do Mat\~ao 1226,\\ Cidade Universit\'aria S\~ao Paulo-SP, 05508-090, Brasil \\
  $^{2}$Facultad de Ciencias Astronómicas y Geofísicas - Universidad Nacional de La~Plata, Paseo del Bosque S/N - 1900 La~Plata, Argentina \\
  $^{3}$ Instituto de Astrofísica de La~Plata (CCT La~Plata - CONICET/UNLP), Argentina \\
  $^{4}$ D´epartement de physique and Centre de Recherche en Astrophysique du Qu\'ebec (CRAQ)\\ Universit\'e de Montr\'eal, C.P. 6128, Succ. Centre-Ville, Montr\'eal, Qu\'ebec, H3C 3J7, Canada\\
  $^{5}$ Instituto Nacional de Pesquisas Espaciais/MCTIC Avenida dos Astronautas 1758, S\~ao Jos\'e dos Campos, SP, 12227-010, Brazil\\
  $^{6}$ Universidade Federal do Rio Grande do Norte, UFRN, Departamento de F\'isica, CP 1641, Natal, RN, 59072-970, Brazil\\
  $^{7}$ CRESST II \& X-ray Astrophysics Laboratory, Code 662, NASA Goddard Space Flight Center, Greenbelt, MD 20771, USA\\
  $^{8}$ Institute for Astrophysics and Computational Sciences, Department of Physics \\
  The Catholic University of America, Washington, DC 20064, USA\\
  $^{9}$ Laboratory for Extraterrestrial Planets and Stellar Astrophysics, Code 667, NASA Goddard Space Flight Center, Greenbelt, MD 20771, USA,\\
  $^{10}$ Department of Physics and Astronomy \& Pittsburgh Particle Physics, Astrophysics, and Cosmology Center (PITT PACC),\\
  $^{11}$San Jos\'e State University, Department of Physics and Astronomy, One Washington Square, San Jos\'e, CA 95192-0106, USA,\\
  $^{12}$ Ritter Observatory, Department of Physics and Astronomy, The University of Toledo, Toledo, OH 43606-3390, USA\\
  $^{13}$ Max Planck Institute for Radio Astronomy, Auf dem H\"{u}gel 69, D-53121 Bonn, Germany\\
$^{14}$Cell Biology and Molecular Genetics Department, University of Maryland, College Park, Maryland 20742-5815, USA \\
}

\date{AUGUST 27th 2018}

\pagerange{\pageref{firstpage}-\pageref{lastpage}} \pubyear{2016}

\def\LaTeX{L\kern-.36em\raise.3ex\hbox{a}\kern-.15em
    T\kern-.1667em\lower.7ex\hbox{E}\kern-.125emX}

\begin{document}

\label{firstpage}

\maketitle
\begin{abstract}
\textcolor{black}{The interacting binary Eta Carinae remains one of the most enigmatic massive stars in our Galaxy despite over four centuries of observations. In this work, its light curve from the ultraviolet to the near-infrared is analysed using spatially resolved HST observations and intense monitoring at the La\,Plata Observatory, combined with previously published photometry. We have developed a method to separate the central stellar object in the ground-based images using HST photometry and applying it to the more numerous ground-based data, which supports the hypothesis that the central source is brightening faster than the almost-constant Homunculus.
After detrending from long-term brightening, the light curve shows periodic orbital modulation ($\Delta V$\,$\sim$\,0.6\,mag) attributed to the wind-wind collision cavity as it sweeps around the primary star and it shows variable projected area to our line-of-sight. Two quasi-periodic components with time scales of 2-3 and 8-10\,yr and low amplitude, $\Delta V$\,$<$\,0.2\,mag, are superimposed on the brightening light-curve, being the only stellar component of variability found, which indicates minimal stellar instability. Moreover, the light curve analysis shows no evidence of ``shell ejections'' at periastron. We propose that the long-term brightening of the stellar core is due to the dissipation of a dusty clump in front of the central star, which works like a natural coronagraph. Thus, the central stars appear to be more stable than previously thought since the dominant variability originates from a changing circumstellar medium. We predict that the brightening phase, due mainly to dust dissipation, will be completed around 2032\,$\pm$\,4\,yr, when the star will be brighter than in the 1600's by up to $\Delta V$\,$\sim$\,1\,mag.
}
\end{abstract}


\newpage

\section{Introduction}
\label{sectionintroduction}
Eta Carinae (\ec) is one of the most enigmatic massive stars in our Galaxy, being frequently observed but not well understood. Estimates of its brightness go back to the 1600's, when it was a relatively faint naked-eye object \citep[$V$\,=\,3.5\,$\pm$\,0.5\,mag;][]{Frew2004}. The star probably started to brighten in the early 1700's, but there were few recorded observations at that time. More frequent observations were prompted by the report of the naturalist William Burchell, who, on a visit to S\~ao Paulo, Brazil in 1827, was surprised to note that the star was very bright ($V$\,=\,0.8-1.5\,mag). Burchell's report helped alert John Herschel, who then monitored the unusual brightness variations from Cape Town (ZA). 

\textcolor{black}{ 
In 1847 \citep{Smith2017} 
there was a large erratic outburst now known as ``The Great Eruption'', whose cause is still unknown. Suggested mechanisms have included stellar mergers \citep{Portegies+2016, Smith+2018b}, super-Eddington eruptions \citep{Owocki2016}, binary interactions at periastron \citep{Kashi2009, Smith+2011}, and pulsational pair instabilities in a massive star \citep{Woosley2017}. 
The kinetic energy of the Homunculus is consistent with an ejection by a faint supernova explosion, not a massive stellar wind \citep{Smith+2003}. Using light-echoes reflected in the Carina nebula in the direction of the Homunculus equator, \citet{Smith+2018a} 
argued for a stellar merger in a triple system as the cause of the eruption.
This appears to account for the kinetic energy, the luminosity burst, the bipolar shape of the Homunculus, and the two successive stages of the velocity field during the Great Eruption.} 

\begin{figure}
 \centering
\includegraphics[width=\linewidth, viewport=10bp 15bp 680bp 530bp]{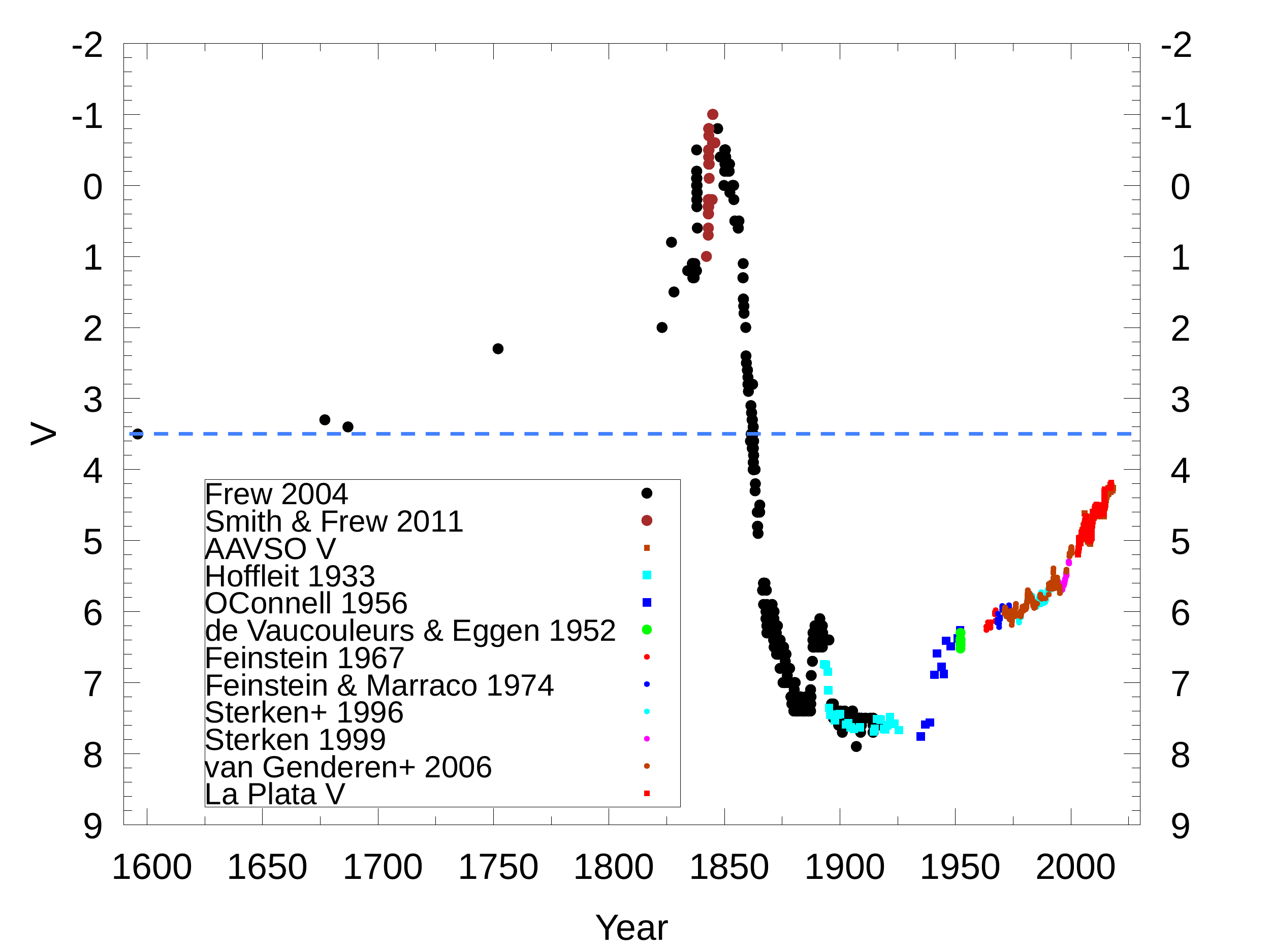}
\caption{Historical light curve of \ec\ from 1600 to 2017 compiled by \citet{Lajus+2009}, with updates from the La\,Plata monitoring campaign, {\tt AAVSO}, and reports from \citet{Frew2004} and \citet{Smith+2011}. The horizontal dashed line indicates the presumed outburst level.}
\label{figlc}
\end{figure}

Figure\,\ref{figlc} collects photometric visual brightness measures of {\ec} (i.e., including the nebula) over more than four centuries. The historical visual estimates, extending from 1592 to 1916, were collected and revised by \citet{Frew2004} and \citet{Smith+2011}. We complemented those data by re-compiling the photographic measurements beginning in 1895 \citep{Hoffleit1933,OConnell1956} with adjustment to fit contemporary visual observations. In the present study we used all available data, plus additional photometric data compiled by \citet{Lajus+2009}, with updates from the La\,Plata monitoring campaign from \citet{Frew2004,Smith+2011} and  {\tt AAVSO}\footnote{We gratefully acknowledge the variable star observations from the {\tt AAVSO} International Database.}. 

After the Great Eruption, the star dramatically faded by over six magnitudes in less than a decade due to the formation of dust in the ejecta,  reaching $V$\,$\sim$\,7.5 by 1880. {\ec} brightened again to $V$\,$\sim$\,6.2 from 1887-1895 in the so-called ``Lesser Eruption'', which is similar to a Luminous Blue Variable (LBV) eruption where  brightness increases by a few magnitudes due to the formation of an expanding opaque, cool pseudo-atmosphere \citep{Humphreys+1999, Kochanek2011}. Kinematic studies of inner structures near the star known as the ``Little-Homunculus'' \citep{Ishibashi+2003} and the ``Weigelt knots'' \citep{Weigelt+1986} indicate that they were ejected during this Lesser Eruption \citep{Weigelt+1995, Smith+2004b}. 

 After the Lesser Eruption, {\ec} again faded to $V$\,$\sim$\,8 where it remained until 1941, when the system suddenly brightened by about one magnitude. In April 1944, \cite{Gaviola1953} recorded the first appearance of high-excitation lines, such as [Ne\,{\sc{iii}}] ($\lambda$\,=\,3868\,{\AA}) and [N\,{\sc{ii}}] ($\lambda$\,=\,5754\,{\AA}), indicating clearing of the dense, internal envelope created by the Lesser Eruption, enabling UV radiation to excite the inner nebular regions. On the other hand, 
 \cite{Abraham+2014} suggested that the 1941-45 brightening was due to a mass ejection, claiming the formation of another nebula they called the ``baby Homunculus''.  This seems unlikely, since mass ejection would add obscuring material causing an accompanying decrease in nebular excitation, which is inconsistent with the presence of high-excitation lines.
 
The high excitation lines noted by \citet{Gaviola1953} were found to be variable by \citet{Zanella+1984}, while \citet{Damineli1996} realised that the sudden disappearance of these lines re-occurred in a 5.5-yr period. 
Independent observations in the X-ray \citep{Corcoran+1995} and radio \citep{Duncan+1995} bands near a low excitation interval in 1992 showed variations correlated with the change in nebular excitation. Subsequent monitoring in the radio, optical, UV, and X-ray regions have now established the system as a massive, long-period, highly eccentric binary \citep{Damineli+1997} in which the observed periodic variations are driven by the collision of the stellar winds of the component stars.
 
After the brightening of the system in 1941, the brightness of \ec\ has increased secularly at an average rate of $\Delta V$\,=\,$-$0.02\,mag\,yr$^{-1}$ up to the { 1990's}\footnote{Since 1967, \ec\ has been the target of considerable photometric monitoring, with different instruments, photometric bands ({\it UBVRI}, Walraven {\it VBLUW}, Geneva {\it UBV$B_1$$B_2$$V_1$G} and Str\"omgren {\it uvby}), photomultipliers, apertures sizes (e.g., \citealt{Feinstein1967, Feinstein+1974, vanGenderen+1984,vanGenderen+1994, vanGenderen+1995, Sterken+1996, vanGenderen+2006}), and CCDs \citep{Sterken+1999, Lajus+2009}}. This slow brightening trend might be produced by a reduction in the total optical depth of obscuring material caused by expansion of the Homunculus, the Little Homunculus and structures close to \ec. Alternatively, it might be due to dust destruction caused by the UV radiation field, and/or an intrinsic brightening of the binary itself.

Starting in the 1990's, observations with the Hubble Space Telescope (HST) have had a revolutionary impact on our knowledge of \ec, resolving the system on spatial scales of $<$\,0\farcs1 (230\,AU).  HST direct imaging and spatially-resolved spectroscopy revealed complex structures with temporal changes which could not be seen in ground-based observations.
HST also enabled imagery into the near ultraviolet (NUV) and spectro-imagery further into the UV down to Ly\,$\alpha$.
For example, it became clear that the narrow forbidden lines are confined to  the Weigelt clumps and other fainter clumpy structures \citep{Davidson+1995, Davidson+1997, Smith2002,Hartman+2005, Nielsen+2005, Nielsen+2007, Gull+2009, Mehner+2010,Richardson+2016, Zethson+2012, Gull+2016}.   Multiple, spatially-resolved complex spectral structures were noted both in the central source \citep{Hillier+2006} and nearby structures including the Weigelt clumps and the fossil winds \citep{Gull+2009, Gull+2016}.     

	From a study of the Homunculus spectrum \citet{HillierandAllen1992} concluded that the extinction to the core along our line of sight was much larger than in other directions. Ground based spectra of the core showed a hybrid spectrum - broad lines ($FWHM$\,$\sim$\,500\,\kms) of  H\,{\sc{i}}, He{\sc{i}}, Fe\,{\sc{ii}} and [Fe\,{\sc{ii}}] together with much narrower ($FWHM$ \,$\sim$\,40\,\kms) lines
of the same species. Conversely, the Homunculus reflected spectrum showed primarily H\,{\sc{i}}, He\,{\sc{i}} and Fe\,{\sc{ii}} lines, with smaller equivalent widths, and the [Fe\,{\sc{ii}}] lines were at least a factor of 5 weaker than in the central star. 
The reflected spectrum showed striking similarities to that of the P-Cygni star HD\,316285. Later \citep{Hillier+2001b} invoked enhanced extinction to explain the presence of the broad [Fe\,{\sc{ii}}] lines in ground-based spectra - they were wind features whose EW was enhanced due to increased attenuation of the continuum by dust along our sight line. The same idea also explains the anomalous strength of the narrow lines that arise from the Weigelt clumps and the discrepant EW of H\,$\alpha$ in direct and nebular reflected spectra. 
The presence of a natural ``coronagraph", like suggested by several authors, can also explain several other observations.

By the end of the 1990's the brightness of the central star as measured with HST was shown to be increasing more rapidly than at earlier epochs \citep{Davidson+1999,Smith+2000,Martin+2006}. Comparison of HST photometry with ground-based observations also showed that the brightness of the central object (the core) was increasing faster than the nebula (nebula = Homunculus + faint inner ejecta). \citet{Hillier+2006} reported slow variation of the ``Weigelt\,D'' clump. The EW of H\,$\alpha$ has declined \citep{Mehner+2012} while higher members of the Balmer emission-line series (which are formed closer to the primary's photosphere) show almost no variability during the last four cycles, as compared to the factor of five increase in the continuum flux (see Figure\,9 in \citealt{Teodoro+2012} for the H\,$\delta$ line). Other features originating in the central core are also fairly stable. For example, the He\,{\sc{ii}} 4686\,{\AA} equivalent width has been repeatable during the last four cycles, despite large variations in the continuum brightness  (see Figure 13 in \citealt{Teodoro+2016}). Constant equivalent width occurs when both the line and the continuum changes by the same factor. Reports by other authors \citep{Davidson+1995,Weigelt+1995,Hillier+2001,Gull+2009,Mehner+2010,Martin+2006} suggested a dissipating  dusty coronagraph in the line-of-sight as the source of the secular brightening.  Polarization measurements on speckle images by \citet{Falcke+1996} indicate that a bar  crosses the central core to the NE direction, which they interpreted as an equatorial disk. They suggest that the B-D Weigelt clumps are part of such a disk. 

Of special interest are the \textit{JHKL} light curves collected over a span of 45 years at the SAA Observatory 
\citep{Whitelock+1983,Whitelock+1994,Whitelock+2004,Feast+2001,Mehner+2014}. Due to the reduced effects of extinction at these NIR wavelengths, the light curve reflects the ionised plasma in the inner system. \citet{Mehner+2014} reported long-term variations of the NIR colour indices, showing that they shift blueward periodically over at least the last four orbital cycles. They interpreted this as an increase in temperature of the stellar wind.  However, other photometric oscillations on various timescales and with various amplitudes have been seen. Some of these variations have been attributed to S\,Doradus (S\,Dor) variability \citep{vanGenderen+1999}, but the expected spectral variability usually associated with S\,Dor was not seen.  
\citet{Whitelock+1994}  
found quasi-periodic oscillations in the \textit{JHKL}-bands on timescales of 1830-1890\,d and 3940\,d in the $V$-band. 
The period length changed from filter to filter. Strictly periodic variability of  P\,=\,5.5\,yr 
was 
recognised 
\citep{Damineli1996,Damineli+1997,Damineli+2000,Damineli+2008a,Teodoro+2016, Corcoran+2017}. 
After the period was established, an inconspicuous 
peak in the visual and NIR  photometry was later shown to be strictly periodic 
\citep{Lajus+2010,Whitelock+2004,Feast+2001}. These periodic peaks have yet to be explained.

In addition to the orbital-related periodicity of the short-lived peaks, 
\citet{vanGenderen+1994} and \citet{Sterken+1996} found low amplitude
(milli-magnitude) variations with a period $P$\,=\,58.58\,d. 
Two decades later, \citet{Richardson+2018} confirmed this period using measurements made with the {\it BRITE}-Constellation nano-satellites. Thus the 58.8\,day variability seems to be stable in frequency over at least four decades. In this paper, 
our focus is to separate the core stellar object from the nebular component and measure their temporal changes.
We follow the convention that photometry of \ec\ includes the bright nebulosity (the Homunculus) within a 10\arcsec radius aperture. 
The paper is organized as follows:
the data and  reduction procedures and methods are described in section \ref{sectionobservations};
Section \ref{sectionresults} discusses the resulting photometry and light curves of the different parts of \ec; Section\,\ref{sectionperiodicity} describes the periodic and quasi-periodic structures in the light curve; Section\,\ref{sectionredlaw} deals with determination of the reddening law and extinction; Section\,\ref{sectiondiscussion}  presents a general view of our results including our predictions for the near future, and finally, Section\,\ref{sectionconclusions} summarises our results.

In the following, we assume the binary system is oriented such that the secondary star is behind the primary at periastron \citep{Gull+2009,Madura12}. This orientation is in agreement with \textcolor{black}{basic observational data, such as: the disappearance at periastron of the high excitation lines in the Weigelt clumps, which are on our side of the binary; the long duration of the low excitation state after periastron; the simultaneous three-dimensional fit of $\rm{He\,\textsc{ii}}$ 4686\,{\AA} seen directly and reflected at the Homunculus South polar cap \citep{Teodoro+2016,Hamaguchi+2014,Corcoran+2017} 
and also from interferometry structures connected to the wind-wind collision at spatial scales of $\approx$\,6\,mas 
\citep[14\,AU;][]{Weigelt+2016}.}

\section{Observations, data reduction and photometric modelling}\label{sectionobservations}
 
\begin{figure}
 \centering
 \includegraphics[width=\linewidth, viewport=10bp 5bp 630bp 540bp]{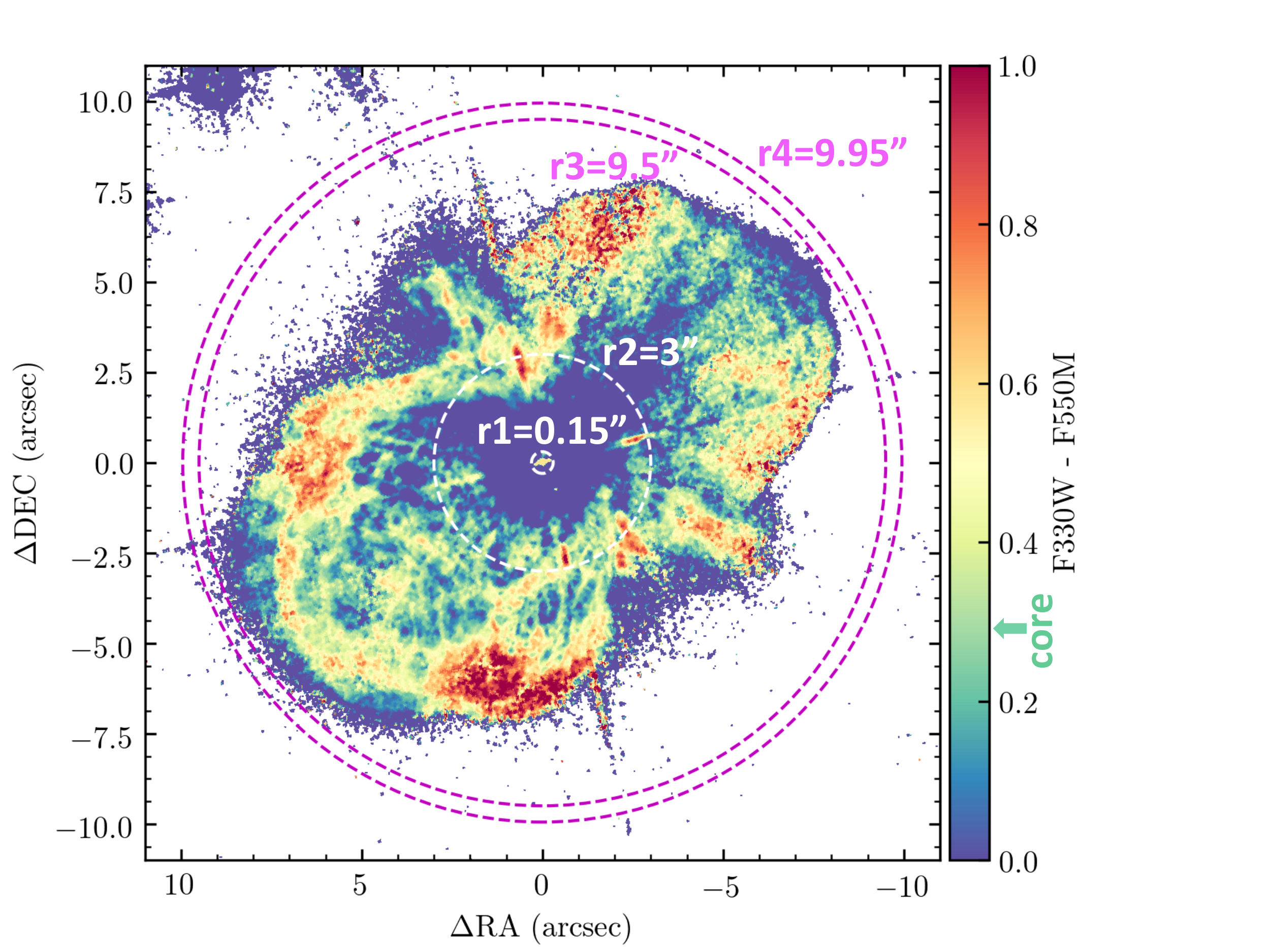}
 \caption{\textcolor{black}{$\rm{F330W}-\rm{F550M}$ colour index map of \ec\ using HST ACS/HRC images taken on 2004 December 06.  
See Section~\ref{aperture} for the definitions of the various apertures.
The central core is redder than the central regions of the nebula, and is bluer than the external parts. The  arrow  at the right of the Figure shows the  colour index of the core as compared to other regions in the colour bar. }}
 \label{figcolormap}
\end{figure}

We used ground-based images from the La\,Plata long-term photometric monitoring campaign\footnote{\url{http://etacar.fcaglp.unlp.edu.ar}} \citep{Lajus+2015} as the primary resource for our data because it provides uniformity of coverage over a long time interval. This is in contrast with the much higher-resolution ACS/HRC images, which while useful to calibrate our ground-based images, are much less frequent and thus less useful to characterise the long-term behaviour of \ec's light curve. We calibrated the brightness of the core in the ground-based images using coeval ACS/HRC images in similar wavebands, in which the star is separated from the nebula. A single ACS/HRC image is sufficient to calibrate all ground-based images. If there are many images, as in the $V$-band, we can take advantage of multiple images to better constrain the zero point.  We split both sets of images into homologous sub-structures, as defined in Figure\,\ref{figcolormap} (see details in sub-section \ref{sectionobservations.3}).

\subsection{Aperture Photometry}
\label{aperture}
We obtained photometry using four different apertures: 
\begin{enumerate}
\item The ``inner aperture'' with radius $r1$\,=\,0\farcs15 centred on the binary.  The inner aperture  is dominated by emission from the binary but includes a faint circumstellar contribution and the wings of the point spread function (PSF) spill outside $r1$. 
\item The ``intermediate aperture'' with radius $r2$\,=\,3\arcsec centred on the binary. 
\item The ``outer aperture'' with radius $r3$\,=\,9\farcs5. (=\,\ec\ after corrected for sky background).  This aperture includes the bulk of the emission from the Homunculus and the binary.
\item The ``external aperture'' with radius $r4$\,=\,9\farcs95.  

\end{enumerate}

We define the ``core'' brightness from the HST imaging photometry as the brightness inside  the inner aperture ($r$\,$\le$\,$r1$) after correcting for loss of light due to the HST point spread function. The net flux measured inside 
the outer aperture ($r$\,$<$\,$r3$) we call the \ec\ flux,  
which contains  the central binary, the Homunculus, and 
a fraction of the surrounding outer nebulosity
\citep[see][for details.]{Walborn+1976, Kiminki+2016}. We used the annulus or ring between the external aperture and the outer aperture ($r3$\,$<$\,$r$\,$\le$\,$r4$) as a measure of the 
background. There are no stars inside the background annulus, and, while there is nebular emission in this region, 
the nebula shows mainly line emission (mostly H\,$\alpha$) which falls outside the broadband filters. 
 
We used tools provided by {\tt IRAF}\footnote{IRAF is distributed by the National Optical Astronomy Observatory, which is operated by the Association of Universities for Research in Astronomy (AURA) under a cooperative agreement with the National Science Foundation.} for the spectrometric and imaging analysis.
  
\subsection{ACS/HRC imaging photometry}
\label{sectionobservations.1}

We measured the brightness of the core,  \ec\, and the nebula (\ec\ minus  core region) from drizzled images obtained by the Advanced Camera for Surveys/High Resolution Camera (ACS/HRC) on {\tt HST}\footnote{This paper was based on observations made with the NASA/ESA
HST. The Hubble Space Telescope is operated by the Association of Universities for Research in Astronomy, Inc., under NASA contract NAS5-26555.} in the F220W, F250W, F330W, FR459M and F550M filters. The extraction of the fluxes was done for the apertures defined in Figure\,\ref{figcolormap}.

To correct the brightness of the core from the surrounding nebular ejecta that contaminate the inner circle, 
we first tried to measure the stellar point-spread function in the ACS/HRC images. However, other stars in the HRC images are much fainter than the core and so could not provide reliable point spread functions (PSFs). 
Therefore, we decided to use aperture photometry and found that an aperture with a radius of 6\,pixels ($r$\,=\,0\farcs15), similar to that used by previous authors \citep{Martin+2006}, isolates most of the light from the stellar core.  

Figure\,\ref{figcolormap} shows the $\rm{F330W}-\rm{F550M}$ colour index map derived from the ACS/HRC images. The inner regions of \ec\ are much bluer than the outer regions. The core is redder than the nebula that surrounds it, and corresponds to emission from a 5800\,K black body; 
the average colour index temperature of the nebular ejecta is $\sim$\,5000\,K and the blue spots in Figure\,\ref{figcolormap} have colours typical of an A-type star.
There are bluer regions close to the central star 
(e.g., $\rm{F330W}-\rm{F550M}$\,$<$\,0.2) and redder regions at the borders of the nebula 
(e.g., $\rm{F330W}-\rm{F550M}$\,$<$\,0.6). It is beyond the scope of the present paper to explore the full range of physical mechanisms that could produce the observed nebular colours.

We used the standard star GD71 in drizzled images to model the ACS/HRC PSF in all filters and measured the encircled energy as a function of the aperture radius. Our measures of the encircled energy fraction were within a few percent of those reported by \citet[][see their Table\,9]{Bohlin2016}. The ACS/HRC point spread function causes $\approx$\,20 per cent of the core's flux to fall outside the $r$\,=\,0\farcs15 core region.   

Magnitudes in the STMAG system derived for the core  (corrected for the fraction of encircled energy) are reported in Table\,\ref{tableHST-lc}. Photometric uncertainties were initially derived from photon counting statistics and are typically $\pm$0.001-0.003 for the F550M images. However these underestimate the true errors, since some images have a different PSF and the shadow of the probe, which is a tiny mirror used for image acquisition, covers $<$\,1\% of the image. The impact on the total flux of \ec\ is much smaller than this percentage since the probe obstructs external portions of the nebula. These obstructed regions were patched using interpolations from surrounding nebulosity. We averaged the photometry performed in images collected as close in time as possible (5 images collected in the period 2003.7-2003.9) to derive the ratio of the mean deviation to individual uncertainty 
measured from photon statistics alone (8$\times$ for the F550M images) for this particular set of images. To define the uncertainty we multiplied the photon statistical error of every image by this quantity. This was done for all filters (F550M, F330W, F250W and F220W). Typical factors are in the range from 2-10. Since \ec\ might have varied in the period when the photometry was averaged, this procedure may result in slightly 
overestimated errors. The typical error was $\pm$\,0.015\,mag for the F550M filter. Our errors are slightly larger than those reported by \citet{Martin+2006} -- their Table\,3 -- for the photometry of the core from ACS/HRC images we have in common.

To derive the flux of the (whole) nebula, we subtracted the flux of the core (corrected by the encircled energy fraction) from that within the $r$\,=\,9\farcs50 aperture.  By using apertures defined above, we separate the flux measurements of the ACS images into the three regions defined by the circles mentioned above:
a) {\it the central circle} ($r$\,=\,0\farcs15): containing the core of the core PSF plus faint nebulosity;
b) {\it the inner ring} (0\farcs15\,$<$\,$r$\,$<$\,3\arcsec): corresponding to the difference between the intermediate and the central circles and composed of the wings of the stellar PSF plus light from the Homunculus lobes; and c) {\it the outer ring} (3\arcsec\,$<$\,$r$\,$<$\,9\farcs50): measured by subtracting the flux in the intermediate aperture ($r$\,=\,3\arcsec) from that of the outer circle  ($r$\,=\,9\farcs50). The flux in the sky annulus was formally subtracted, but it is very close to zero in  the drizzled ACS/HRC images for all filters.

As an example of the distribution of light in the various apertures
we used an  image recorded through the F550M filter in 2005.  We found that 50\% of the total flux of the \ec\ region ($r$\,$<$\,9\farcs5) is contained within 0\farcs5 of the star, while
80\% of the total flux is contained within 3\arcsec of the star, and the 6.5\arcsec\,$<$\,$r$\,$<$\,9\farcs5 annulus contains just 4\% of the total flux.

\subsection{Synthetic and narrow band  photometry of the core from STIS spectra}
\label{sectionobservations.2}

We used available STIS spectra to derive synthetic photometry of the core region. 
We downloaded all relevant 2D spectra from the {\tt \ec\ Treasury Program}\footnote{http://etacar.umn.edu/} archive and extracted and summed the spectra within an aperture of 5\,pixels in the spatial direction to get one-dimensional spectra. This aperture corresponds to a 0\farcs125\,$\times$\,0\farcs15 rectangle centred on the core.
We defined five wavebands, including the nominal central wavelengths of the ACS/HRC filters, and extracted magnitudes in these defined bands. We applied efficiency factors as a function of wavelength in the same way as applied to the ACS/HRC images \citep{Bohlin2016}, adopting a 0\farcs15 radius aperture. Though this is not the real aperture used in the spectral extraction, it still preserves the relative fluxes along the SED, and the absolute fluxes can be easily derived. We compared the fluxes of the core measured in ACS/HRC contemporary or time-interpolated images with those measured in the spectra to derive the calibration constants. The average values were applied to the narrow-band fluxes to calibrate them in absolute flux units (erg\,cm$^{-2}$\,s$^{-1}$\,\AA$^{-1}$) and also to calibrate the extracted unidimensional spectra. 
 These results were compared to those obtained through the traditional synthetic photometry procedure (that is, folding the spectra with the filter pass-bands), which indicate a good match between the methods.

The advantage of using narrow-bands instead of synthetic photometry is that we can measure fluxes even at the borders of some spectra, which would not be possible with the broad filters, because part of the passband falls outside the spectral wavelength range. Narrow band extraction from STIS spectra taken between 1998.22 and 2004.15 are in excellent agreement with those calibrated by \citet{Hillier+2001} and \citet{Hillier+2006}. 

\textcolor{black}{ We adopted a similar procedure to that used for ACS/HRC photometry to derive errors for the broad-band and narrow-band photometry derived from STIS spectra. The errors in these two techniques are larger than for ACS/HRC photometry because a final STIS spectrum covering a wide wavelength range is the result of a combination of shorter spectra, each one for a particular grating angle. Individual sub-spectra overlap accurately in the central wavelength, but not very well at the borders. The error in the combined STIS spectrum was measured as the average dispersion of the individual sub-spectra, which depends on the wavelength range, 
but in general 
it is 
larger towards the UV.  
 There are at least two spectra observed with the same telescope pointing, and for broad-band synthetic photometry the typical error is $\pm$\,0.06\,mag. In the case of narrow-band photometry, there are 
 numerous sub-spectra covering the same wavelength interval taken over a time scale of a few days. The typical error for photometry done by integrating over the narrow-band STIS spectra is $\pm$\,0.03\,mag.}

\subsection{La\,Plata broad-band imaging photometry}
\label{sectionobservations.3}
Ground-based photometric observations were part of a long-term monitoring campaign begun in Jan 2003. More than 60,000 \textit{BVRI} images were acquired using a CCD camera attached to the 0.8-m \textit{Virpi Niemela} telescope at \emph{La\,Plata Observatory} (Argentina).
Instrumentation and observing methodology, which was the same through the whole campaign,  have been described by \citet{Lajus+2009}. The monitoring continued until the end of Sept 2014 and then resumed again from January through  May 2017. The typical stellar point source has a $FWHM$\,=\,3\arcsec,  due to optics and seeing, with variations  from 1\farcs5-1\farcs8 on times of good seeing to as large as  7\arcsec when seeing conditions were extremely poor. About 10-50 images were taken every night, sampling the seeing in a wide range of conditions.
We extracted brightnesses using aperture photometry with an outer circle of 11\farcs8 to include the Homunculus and exclude the closest stars (Tr16-64, Tr16-65, and Tr16-66). The outer circle in the ground-based images must be larger than used in photometric analysis of the ACS/HRC images to account  for the effects of seeing.
Sky background in the ground-based images was extracted within the annulus: 11\farcs8\,$<$\,$r$\,$<$\,17\farcs76. The field of view (FOV) of ACS images does not allow such a large radius. Nebular knots in the ground-based sky annulus are even fainter than those in HST images.

For the $U$-band we used data collected with the 4.1-m {\tt SOAR}\footnote{Based on observations obtained at the Southern Astrophysical Research ({\tt SOAR}) telescope, which is a joint project of the Minist\'erio da Ci\^encia, Tecnologia, e Inovac\~ao (MCTI) da Rep\'ublica Federativa do Brasil, the U.S. National Optical Astronomy Observatory (NOAO), the University of North Carolina at Chapel Hill (UNC), and Michigan State University (MSU)} telescope in the period between December 2008 and May 2017, and data taken at the 0.6-m {\tt Casleo}\footnote{{\tt CASLEO} is operated under agreement between CONICET, and the Universities of La\,Plata, C\'ordoba, and San Juan, Argentina.} telescope between December 2008 and April 2009. In addition to these data, we used published photometry in $U$, $B$ and $V$ taken before 2003 and after 2014.5 for \ec\ from the {\tt AAVSO} database.
All historical photometry was obtained using similar apertures for \ec, so that differences in sampling the outer faint nebular regions are not larger than those produced by variations in filter pass-bands and photometric errors. 
We shifted the published magnitudes by small amounts (about 0.05-0.20\,mag) to match the La\,Plata light curve.

Our main goal is to extract the light curve of the core from the $V$-band. To do this, the flux inside several apertures in La\,Plata images were extracted and the nebular fluxes were calibrated with ACS/HRC images and subtracted from the whole \ec\ object. Ground-based images are blurred by the seeing, in addition to the PSF of the telescope, which has a smaller aperture than HST. Even so, when using coeval images, we can subtract the flux of the core from the corresponding ground-based images and recover the flux of the nebula. Since we extracted the flux of the outer ring, which measures essentially the external parts of the nebula, we could use it for other nights to calibrate the flux of the entire nebula, as long as it varies much less than the core. This seems to be the case, as shown by Figure \ref{starnebratio}, based on ACS/HRC images. This scheme works safely if we are able to derive relatively frequent calibrations of the (outer ring)/nebula ratio and avoid phases close to periastron passages. But, the question is how to do that in practice, since every ground-based image is blurred by a different seeing and we should derive this ratio for every single image.

As the seeing worsens, light from within the $r$\,$<$\,3\arcsec aperture spills into the  $r$\,$>$\,3\arcsec aperture, while light from the outer annulus (3\arcsec\,$<$\,$r$\,$<$\,11\farcs8) spills into the outer and inner apertures. The amount of light lost from each of these two apertures is a function of the seeing.
Our task is therefore to find such a function and reduce flux measurements to a single number corresponding to a representative seeing value, that characterises the entire night.

\begin{figure}
 \centering
 \includegraphics[width=\linewidth, viewport=35bp 20bp 525bp 515bp]{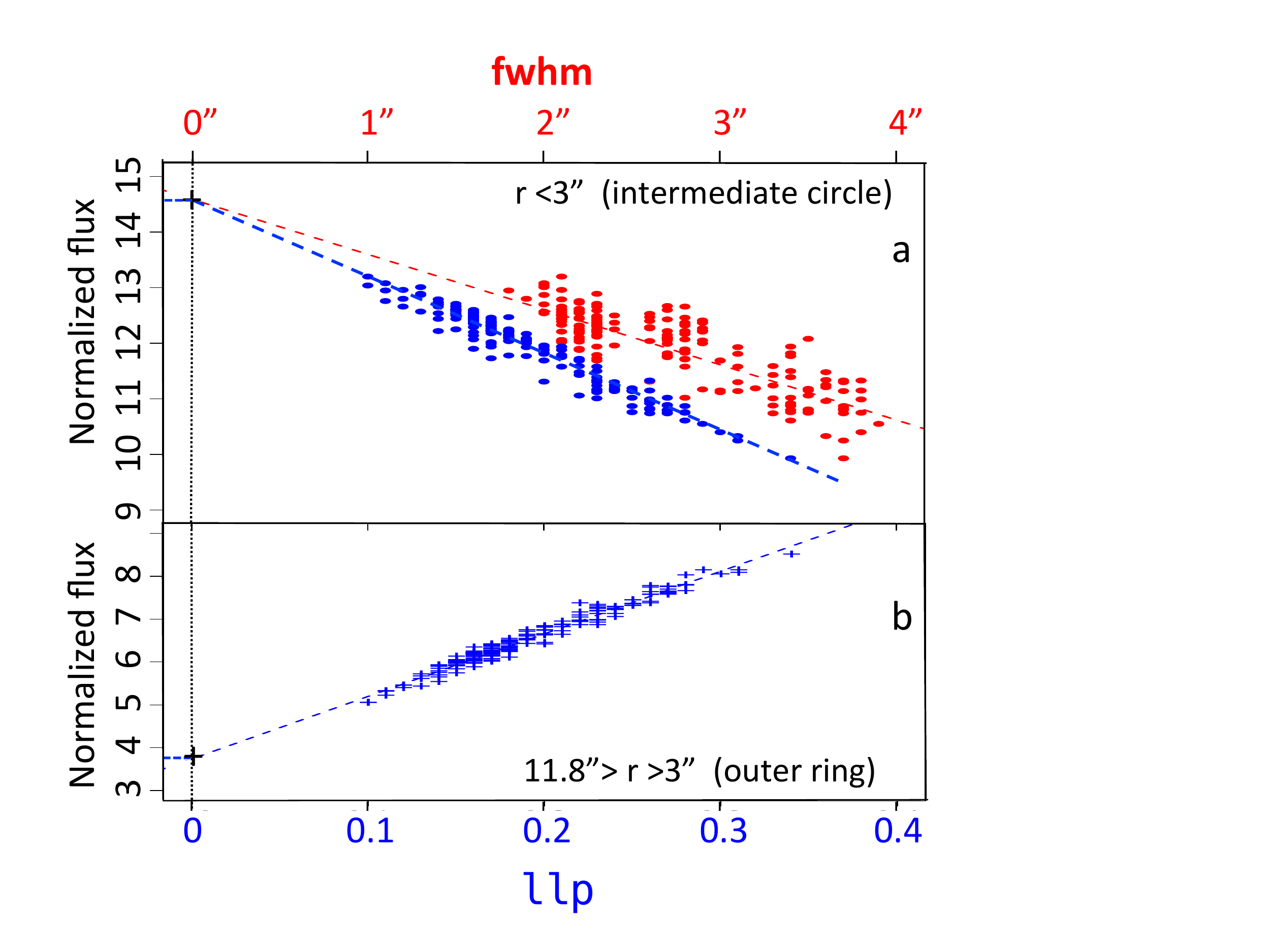}
 \caption{
 \ec\ $V$-band flux inside two apertures as a function of the seeing for a typical La\,Plata night (20 January 2007). The vertical axes show the ratio of the flux of \ec\ within the specified aperture to that of the total flux of the comparison star HDE 303308 (taken within an aperture of $r$\,$<$\,11\farcs8 from the star). \textit{a)} The red dots use the measured $FWHM$ of HDE\,303308 shown by the red labels on the horizontal axis.  
The blue dots are parametrised by the ``light-loss'' (see text) as given by the blue labels on the horizontal axis. The flux of \ec\ inside the intermediate aperture   decreases as the seeing increases.
\textit{b)}
The variation of the flux ratio for the 3\arcsec\,$<$\,$r$\,$<$\,11\farcs8 annulus parametrised by the light-loss.  
The results of each night are characterized by the average of the \ec\ normalised fluxes and by the intercept of the linear fit through the outer ring fluxes (at $LLP$\,=\,0). The flux inside the intermediate circle at every seeing parameter is the difference between these two. Typical formal errors for the extrapolated $V$-band values are about 2$\times$ larger than the size of the symbols in the plot. 
}
 \label{figlightloss}
\end{figure}

We first used the full width at half maximum ($FWHM$) of the point spread function (PSF) of the comparison star HDE\,303308 (located 1\arcmin to the north of \ec) for a measure of the seeing. In the upper panel of Figure \ref{figlightloss} we plot the ratio of the flux of \ec\ in the intermediate circle aperture to that of the comparison star (within an aperture of $r$\,$<$\,11\farcs8 of HDE\,303308 to capture the total stellar flux) as a function of $FWHM$ 
for a representative night (20 January 2007 in this example).  We found that the empirical function turned out to be a simple linear relation for $FWHM$ larger than $\sim$2\arcsec (``bad seeing regime''). For more details, see Appendix\,\ref{appendix_b}. In this way, the flux at the intercept of the linear fit ($FWHM$\,$=$\,0) is a natural representation of the entire night. The rare images in the La\,Plata data bank which have seeing better than the ``bad seeing regime'' are automatically skipped in the fitting process. In this way, all measurements in every night were reduced to two numbers:
$a)$ the average flux in the outer circle; and
$b)$ the intercept of the linear fit to the outer ring fluxes.
The intermediate circle is a dependent parameter: it is just the difference between the flux of \ec\ (outer circle) and that of the outer ring.

	Although the seeing primarily affects the $FWHM$ of the source image, it also affects the wings of the PSF and has non-symmetric components. We then attempted  a more robust characterization of the seeing, which we call the \textit{light loss} parameter ($LLP$).  The $LLP$ is defined as the fraction of flux from a point source spilled outside a defined circle, normalised by the total flux of the source as measured inside an aperture sufficiently large as to contain the whole flux. Its relation with the $FWHM$ is shown in Figure \ref{llpxfwhm}. The $LLP$  was measured using  HDE\,303308 as a comparison star and defined as:
\begin{equation}
LLP_{\rm HDE\,303308} = \frac{f11\farcs8 - f3\farcs0}{f11\farcs8}
\label{eqllf}
\end{equation}

\noindent
\textcolor{black}{
where $LLP$ is the \textit{light loss} parameter measured for  HDE\,303308; $f11\farcs8$ is the total flux inside $r$\,=\,11\farcs8, which is seeing-independent; and $f3\farcs0$ is the seeing-dependent flux inside $r$\,=\,3\farcs0. These aperture fluxes are normalized to that measured inside  $r$\,=\,11\farcs8 in the comparison star, in order to take advantage of the higher accuracy using differential photometry.}

	Regarding \ec, fluxes inside the intermediate  aperture ($r$\,=\,3\farcs0) are shown as blue dots in panel (a) of Figure\,\ref{figlightloss} for which the $x$-axis values are indicated as $LLP$ in blue colour (ranging from 0 to 0.4). Although the flux at the intercept is the same as  that derived using the $FWHM$ as the seeing parameter, the correlation of the extracted flux with $LLP$ is better than the correlation with $FWHM$. This is relevant for nights when the seeing is not sampled over a wide range of seeing conditions. We chose $r$\,=\,3\arcsec to define the \textit{light loss} parameter because it is about 2$\times$ the median seeing and $LLP$ is not too sensitive to larger apertures in a  number of nights we have examined.
The bottom panel in Figure\,\ref{figlightloss} shows the flux ratio inside the outer ring annulus aperture of \ec\ as a function of the $LLP$ parameter. Fluxes at the intercept correspond to the measurement in the intermediate circle and outer ring apertures  divided by the flux of HDE\,303308. Magnitudes of HDE\,303308 are $V$\,=\,8.15\,mag and $B$\,=\,8.27\,mag \citep{Feinstein1982}.

\section{Results}
\label{sectionresults}

\begin{figure}
 \centering
 \resizebox{\hsize}{!}{\includegraphics[width=\linewidth, viewport=0bp 0bp 590bp 520bp]{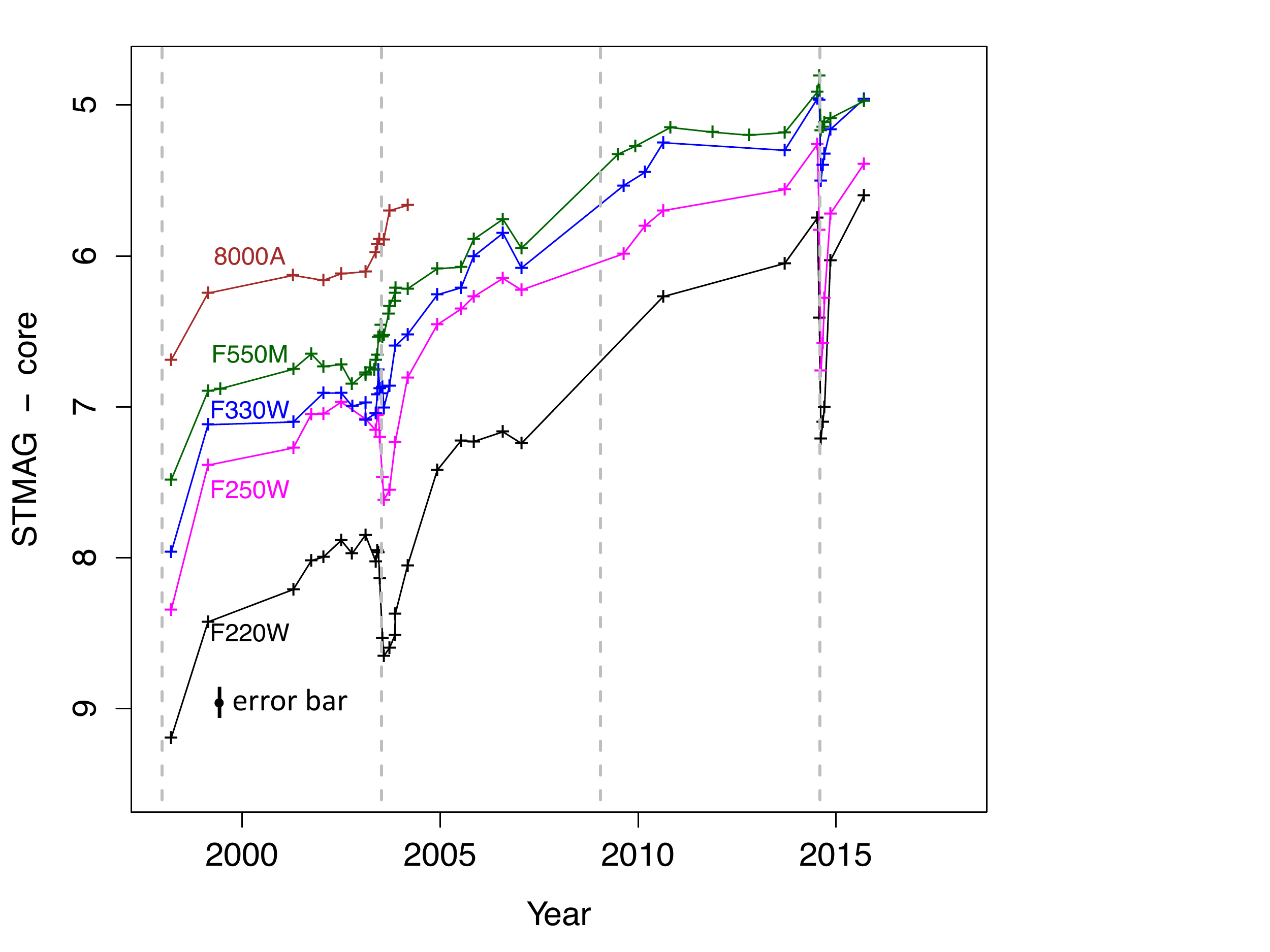}} \\[-3.0ex]
 \caption{Magnitudes in the STMAG system of the core measured in five spectral bands: F220W, F250W, F330W, F550M and a narrow band centred on 8000\,{\AA}, derived from STIS spectra. Vertical dashed lines indicate times of periastron passages as determined by \citet{Teodoro+2016}. We combined ACS/HRC flux measures with brightnesses derived from STIS synthetic photometry. Long-term brightening is clearly seen in all filters. The light curves have different behaviours at periastron passage. In particular, for wavelengths shorter than 3363\,{\AA} (F330W) there is a deep, long post-periastron minimum which is not very apparent at longer wavelengths. Note: No observations were conducted with HST across the 2009 periastron event. Typical error bars are $\pm$\,0.015\,mag for ACS/HRC and $\pm$\,0.03\,mag for STIS narrow-bands, and $\pm$\,0.06\,mag  for synthetic photometry. }
 \label{stmag-star} 
\end{figure}

\subsection{HST observations}
\subsubsection{Light curves of the core}
\label{sectionresults.1}

\begin{figure}
 \centering
 \resizebox{\hsize}{!}{\includegraphics[width=\linewidth, viewport=0bp 0bp 580bp 520bp]{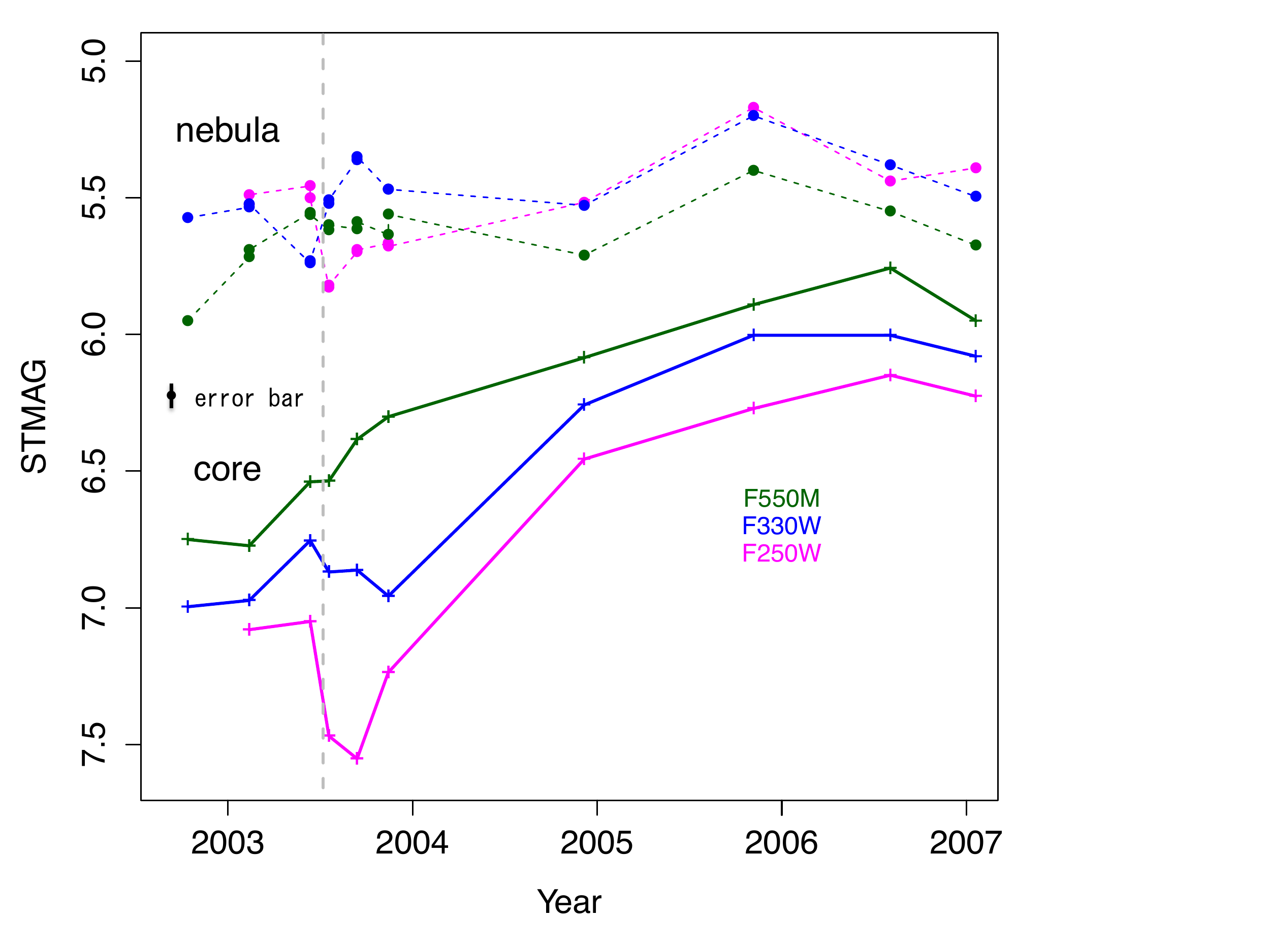}} \\[-3.0ex]
 \caption{Brightening evolution of the core magnitudes ({\it solid lines}) and the nebula ({\it dashed lines}) using ACS/HRC data.Notice that the integrated flux over the nebula varies little across periastron passage, while the core changes substantially. The core brightened faster than the nebula in this time interval, but remained fainter until 2007.0 at all wavelengths shorter than the $V$-band. The nebula displayed a ``local maximum'' around 2006 in all the three filters.}
 \label{starneb} 
\end{figure}

Our photometry for the core is shown in Figure\,\ref{stmag-star}. Magnitudes are in the STMAG system and were extracted: 
$a)$ from ACS/HRC images in F220W, F250W, F330W and F550M filters;
$b)$ from synthetic photometry in STIS spectra using the same pass-bands as these filters; and 
$c)$ from narrow-band photometry in STIS spectra -- see Table\,\ref{tableHST-lc} in the Appendix\,\ref{appendix_a}. At the bottom of that Table, we also give the magnitudes (STMAG system) of the stellar model calculated by \citet{Hillier+2001} for a distance of 2.3\,kpc.  Although the flux increase in the five filters are similar, the slope at shorter wavelengths (e.g., F220W and F250W) is slightly steeper than that at longer wavelengths (e.g., F550M).
Another remarkable feature is the post-periastron minimum, which increases dramatically towards shorter wavelengths and is barely detectable at wavelengths longer than the $V$-band.

The depth and duration of the post-periastron minimum is wavelength-dependent -- it is shallow and short at $\rm{F330W}$, but it is $\sim$1\,mag deep and lasts more than a year at $\rm{F220W}$. The ``brightness jump'' across periastron passage in the optical and NIR light curves claimed in previous work \citep{Mehner+2014} is probably due to the recovery after the periastron dip combined with the overall secular brightening (see Section\,\ref{sectionperiodicity} and Figure\,\ref{MainOscillation}).

Figure\,\ref{starneb} shows the brightness evolution of the core and the nebula from ACS/HRC images. Long-term brightening is clearly seen with fluctuations near periastron passage and near 2007.0, as previously reported by \citet{Martin+2010}.  The nebula shows only small brightness fluctuations at mid-cycle (around 2006.0), 
consistent with ground-based observations (see Figure\,\ref{V-all} ).  The nebular flux has minor fluctuations around periastron passage, as expected for the moving spots close to the central star reported by \citet{Smith+2004a}. 
When ACS/HRC monitoring ceased in 2007.1, the nebula was still brighter than the core at all wavelengths shorter than F550M.

\subsubsection{Core versus nebula light curves}
\label{sectionresults.2}

\begin{figure}
 \centering
\includegraphics[width=\linewidth, viewport=50bp 0bp 650bp 540bp]{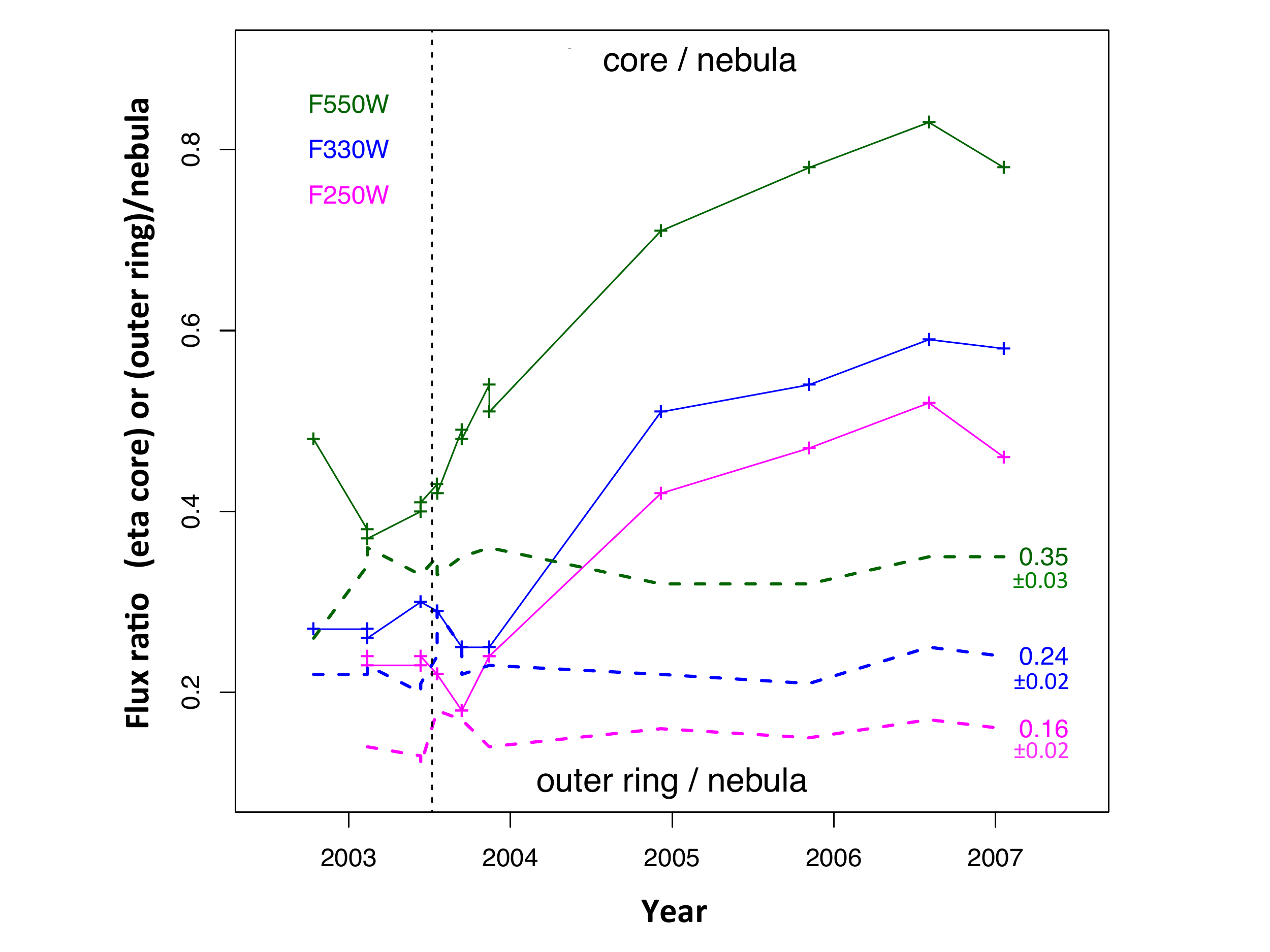} \\[-3.0ex]
 \caption{Evolution of the flux ratios
 $a)$ (core)/(nebula) -- solid lines; and
 $b)$ (outer ring)/(nebula) -- dashed lines. 
 Numbers at the right of the (outer ring)/(nebula) ratios are median values.  Their respective errors are $\sim$10\%. In contrast, the (core)/(nebula) ratios varied at a level $\approx$\,15 times higher.
 The contrast between the core and the nebula increases with the wavelength because the colour of the core is much redder than that of the nebula.
 }
 \label{starnebratio} 
\end{figure}

An important discussion in this paper is the comparison between the brightening of the core, as compared to that of the nebula. In  Figure \ref{starnebratio} the solid lines show that the core brightened twice as much as the nebula in the period 2003-2007. This is in agreement with a number of previous works as discussed in the Introduction, indicating that the nebula is not reflecting the brightening we have been witnessing from our vantage point. That Figure displays an additional parameter: the ratio between the outer ring and the nebula. As shown by dashed lines, the ratio (outer ring)/nebula has not varied more than 10 percent in the same period. Although we know that there are some moving spots in the inner nebula around periastron passages \citep{Smith+2004a}, their integrated magnitudes  are more constant than we anticipated. The fact that this ratio increases from the UV to the optical indicates that the outer parts of the nebula are redder than the inner regions, as otherwise displayed in Figure \ref{figcolormap}.

\subsubsection{Colour index light curves}
\label{sectionresults.3}

\begin{figure}
 \centering
 \resizebox{\hsize}{!}{\includegraphics[width=\linewidth, viewport=0bp 0bp 590bp 520bp]{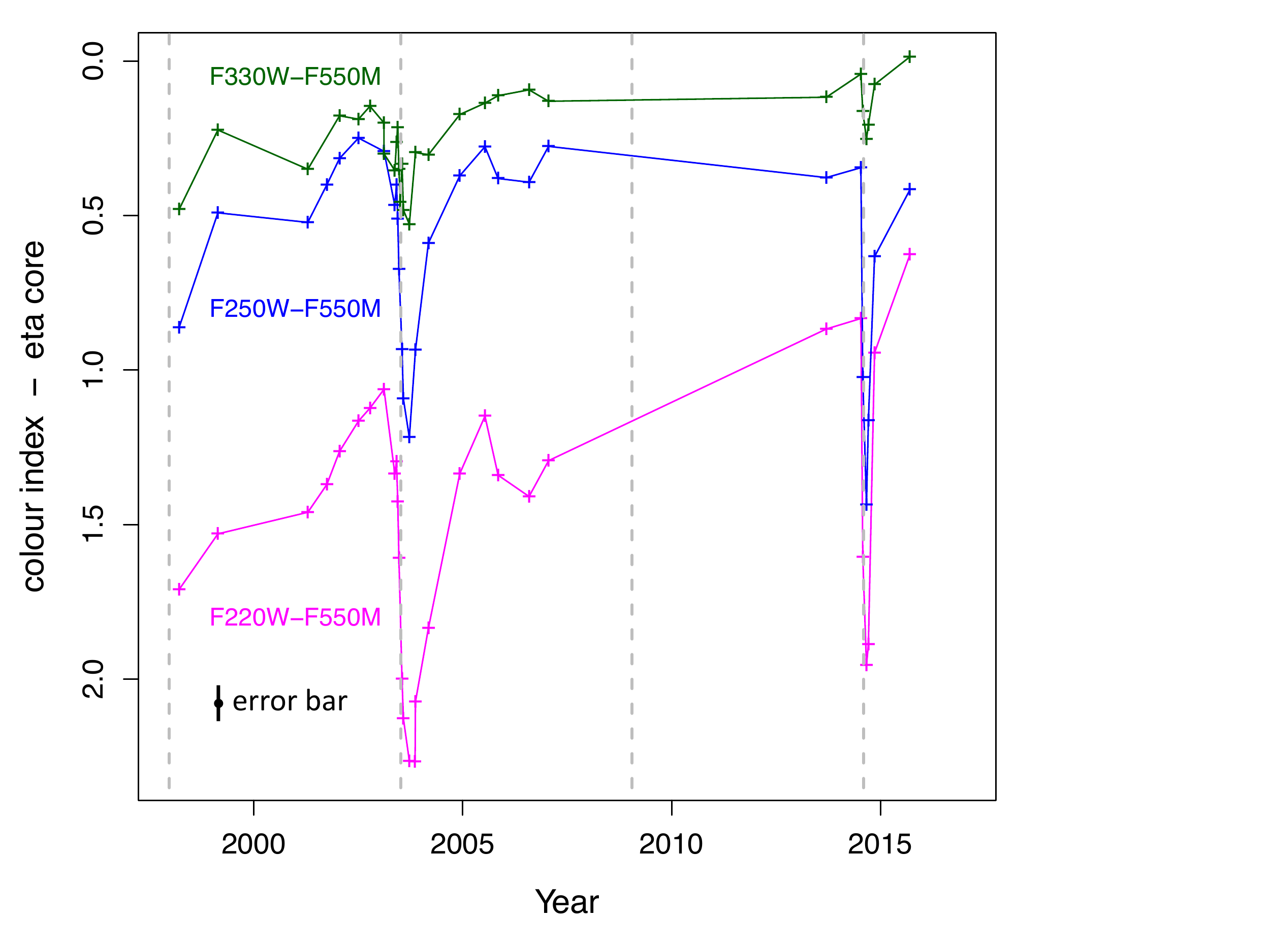}} \\[-3.0ex]
 \caption{Colour indices of the core from HST (ACS plus STIS synthetic photometry). 
 No large decrease in colour index is seen in any pair of filters. The  $\rm{F220W}-\rm{F550M}$ colour index
 shows a faint long-term trend, probably indicating that small grains are not involved in the brightening. The recovery from the brightness dip near periastron (the ``P-dip'') is a function of excitation energy, and takes about a year. This is similar to the behaviour of doubly ionized forbidden lines and suggests a similar mechanism \citep{Damineli+2008b}. Vertical dashed lines mark periastron passages, as in all other light-curve plots. 
 Error bars for the colour indices are typically $\pm$\,0.05\,mag and a little smaller for the sub-set of ACS/HRC images.}
 \label{colormag-star} 
\end{figure}

Figure\,\ref{colormag-star} exhibits the three colour indices defined by the ACS/HRC filters, which we are comparing here. The long-term behaviour shows a blueing, which is larger for the $\rm{F220W}-\rm{F550M}$ index than the other two. This is as expected for dust extinction, but when compared to the large brightening in the same time interval (see Figure\,\ref{stmag-star}), the colour changes are smaller than expected for a typical ISM reddening law ($R_{\rm V}$\,=\,3.1), indicating a larger $R$ value. This suggests that the observed brightening is caused by destruction of grains larger than those in the ISM. The $\rm{F330W}-\rm{F550M}$ colour shows the shallowest post-periastron minimum between the three colours. It decreases by 0.5\,mag soon after periastron passage, as compared to the large decrease of 2.3\,mag in the $\rm{F220W}-\rm{F550M}$ colour index.

\subsection{Ground-based observations}
\subsubsection{Calibration of ground-based photometry using ACS/HRC images}
\label{sectionobservations.1}

\begin{figure}
 \centering
 \resizebox{\hsize}{!}
 {\includegraphics[width=\linewidth, viewport=0bp 0bp 580bp 520bp]{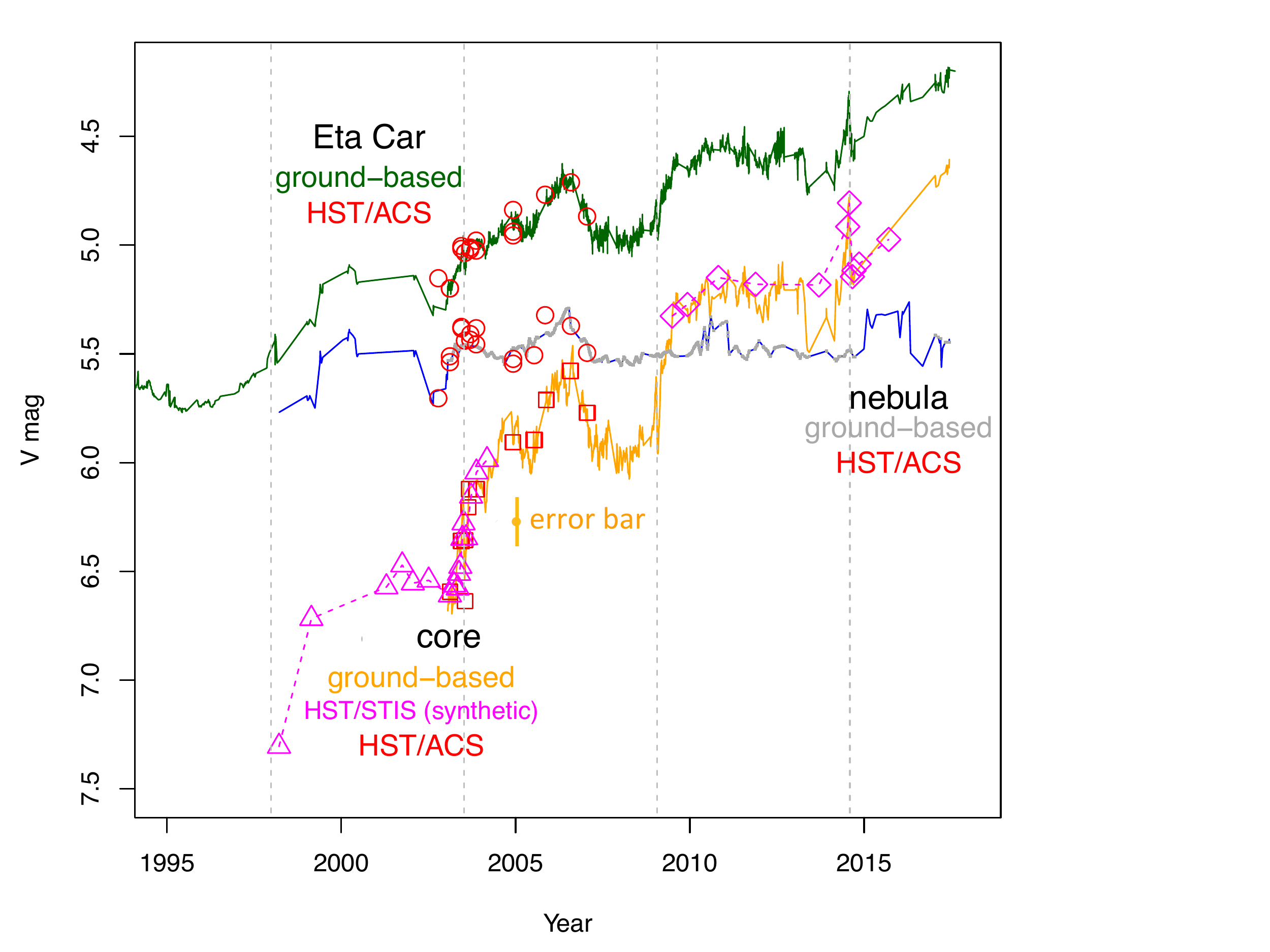}} \\[-3.0ex]
 \caption{Ground-based $V$-band magnitudes of \ec\ and its two components.  {\it Empty red circles} and {\it squares} represent measurements from ACS/HRC F550M images, transformed to the $V$-band ($V$\,=\,F550STMAG\,$-$\,0.18) which were used to calibrate the ground-based core fluxes. {\it Magenta polygons} are from synthetic photometry -- triangles for F550M filter convolution and diamonds for integration in 10\,{\AA} narrow bands centred at 5580\,\AA; {\it green solid line} is the \ec\ historic light curve; {\it grey solid curve} is the light curve of the nebula derived directly from the outer ring, and {\it orange solid line} is the resulting ground-based light curve of the core. The central stellar object became brighter than the nebula in 2010. The {\it blue line} is the nebula light curve as measured by just subtracting the core's fluxes from those of \ec. When doing subtraction between non-coeval measurements we interpolated the light curve. Typical error bars for the average \ec\ ground-based differential photometry each night are $\pm$\,0.05\,mag, and  $\pm$\,0.1\,mag for the extracted stellar core because of systematic uncertainty in the calibration.}
 \label{V-all} 
\end{figure}

To recover the flux of the core from all ground-based images from La\,Plata monitoring, we followed the following steps:
\begin{enumerate}
\item For 15 contemporary ACS/HRC images taken in the interval 2003.1 to 2007.05, we compared the magnitude of \ec\ in the $\rm{F550M}$ filter (ACS/HRC) with that in the $V$-band of La\,Plata (shown as red circles and solid green line, respectively, in Figure\,\ref{V-all}, under the label ``{\ec}'') and derived the transformation constant between these two filters ($V$\,=\,$\rm{F550M(STMAG)}$\,$-$\,$0.18$).
\item We applied the same constant to transform the $\rm{F550M}$ filter flux (ACS/HRC) to $V$ for the core and the nebula (empty red squares and red circles in the zones assigned as ``core'' and ``nebula'', respectively, in Figure\,\ref{V-all}).

\item We subtracted the contribution of the core in the transformed ACS/HRC $V$-band from the \ec\ ground-based $V$-band flux to derive the $V$-band flux of the nebula in the ground-based images for the 15 epochs.

\item The intercept of the linear fit to the outer ring aperture for each night was subtracted from the nebula flux to derive the (outer ring)/(nebula) ratio. The average of the 15 nights results in (outer ring)/(nebula)\,=\,0.37\,$\pm$\,0.04.

\item The small variation of the (outer ring)/nebula ratio in ground based images in a period when the core varied by a large amount ($\Delta V$\,$>$\,1\,mag) is similar to what was found by direct analysis of ACS/HRC images (Figure\,\ref{starnebratio}). We are thus encouraged to apply such a ratio value to the entire set of ground-based data. For every night, the intercept of the outer ring fit was divided by 0.37 to derive the flux of the nebula in the $V$-band. 

\item We subtracted the flux of the nebula from that of  \ec\  to derive the corrected magnitude of the core in all ground-based images, which is shown as the solid orange line in Figure\,\ref{V-all}. Although the core magnitude at every night has a small formal error, we adopt the calibration uncertainty $\pm$\,0.05 as the uncertainty in the core's $V$-band magnitudes.

\end{enumerate}

The 2006-2007.5 ``local maximum'' in the nebula light curve is a genuine peak in the core light curve reflected by the nebula. Although we cannot derive accurate values, the nebula maximum was $\Delta V$\,$\approx$\,0.2\,mag. The corresponding ``local maximum`'' in the core is hinted by the unusual maximum in  of the light-curve and seems to have a larger amplitude than the nebula, but its coincidence with a larger variability -- the orbital modulation, see Section\,\ref{sectionperiodicity} -- prevents an accurate measurement of the local maximum amplitude.

In addition to these ``local maxima'' detected outside periastron there are known  ``spots'' in the inner part of the nebula reported by \citet{Smith+2004a} close to periastron. However, the integrated flux of the nebula is almost constant, since the flux variation by these spots is much lower than that of the entire nebula. Although these features do not impact our analysis, the photometry very close to periastron should be regarded with caution.

Figure\,\ref{V-all} also includes the synthetic photometry in the F550M filter measurements transformed to the $V$-band (shown as magenta triangles). The transformation constant is the same as that obtained for the ACS/HRC images. It can be seen that the core exceeded the brightness of the nebula around 2010.0.  All the photometric and synthetic photometry data were already in place when we decided to perform narrow-band photometry for the recently released STIS spectra for times after 2009.445 (last group of magenta diamonds in Figure\,\ref{V-all}). The good match between these data and those for the core as recovered from the ground-based images (orange line) is very important, as any significant systematic effect would have caused a mismatch between these two groups of points. 
 
 The nebula light curve shows 3-4 peaks $\sim$\,0.1\,mag high. The peak that occurred in 2006.5, with 2\,years duration, was also detected in the ACS/HRC images, which indicates that those variations are probably real.  Figure\,\ref{B-lc} for the $B$-filter  also shows a 0.1\,mag peak for the same date.  Other 10\% high peaks are seen in both $V$- and $B$-band light curves of the nebula around 2011 and 2016, but the absence of corresponding peaks in the core light curve brings doubt about their reality.
 
 	On top of those local maxima the nebula has a long-term brightening, but at a very slow pace. This brightening is not readily apparent in the ACS images because of the limited time span of the HST photometry. The slow brightening of the nebula might be produced by the compensating effects of its expansion and consequent dilution of the dust column density between the Homunculus lobes and the core.

From Figure\,\ref{V-all} it is clear that the lobes of the Homunculus do not experience the same fast brightening of the core as it is seen from our line-of-sight. The core brightening must be produced by an entirely different mechanism from the simple expansion-driven dust dilution that brightens the nebula slowly.

\subsubsection{$B$-band light curve of the core recovered from ground-based images}
\label{subsection.2}
 
We derived calibrated $B$-band photometry of the core region to compare to the changes observed in $V$. 
There are only two ACS/HRC images (taken at the same date) in the $\rm{F459M}$ filter (taken on 2003.9) which could be used to calibrate the $B$-band ground-based photometry. We averaged the core brightness for those two images and followed the procedure given above to derive corrected $B$-band light curves for the core region.

\begin{figure}
 \centering
 \resizebox{\hsize}{!}{\includegraphics[width=\linewidth, viewport=0bp 0bp 580bp 520bp]{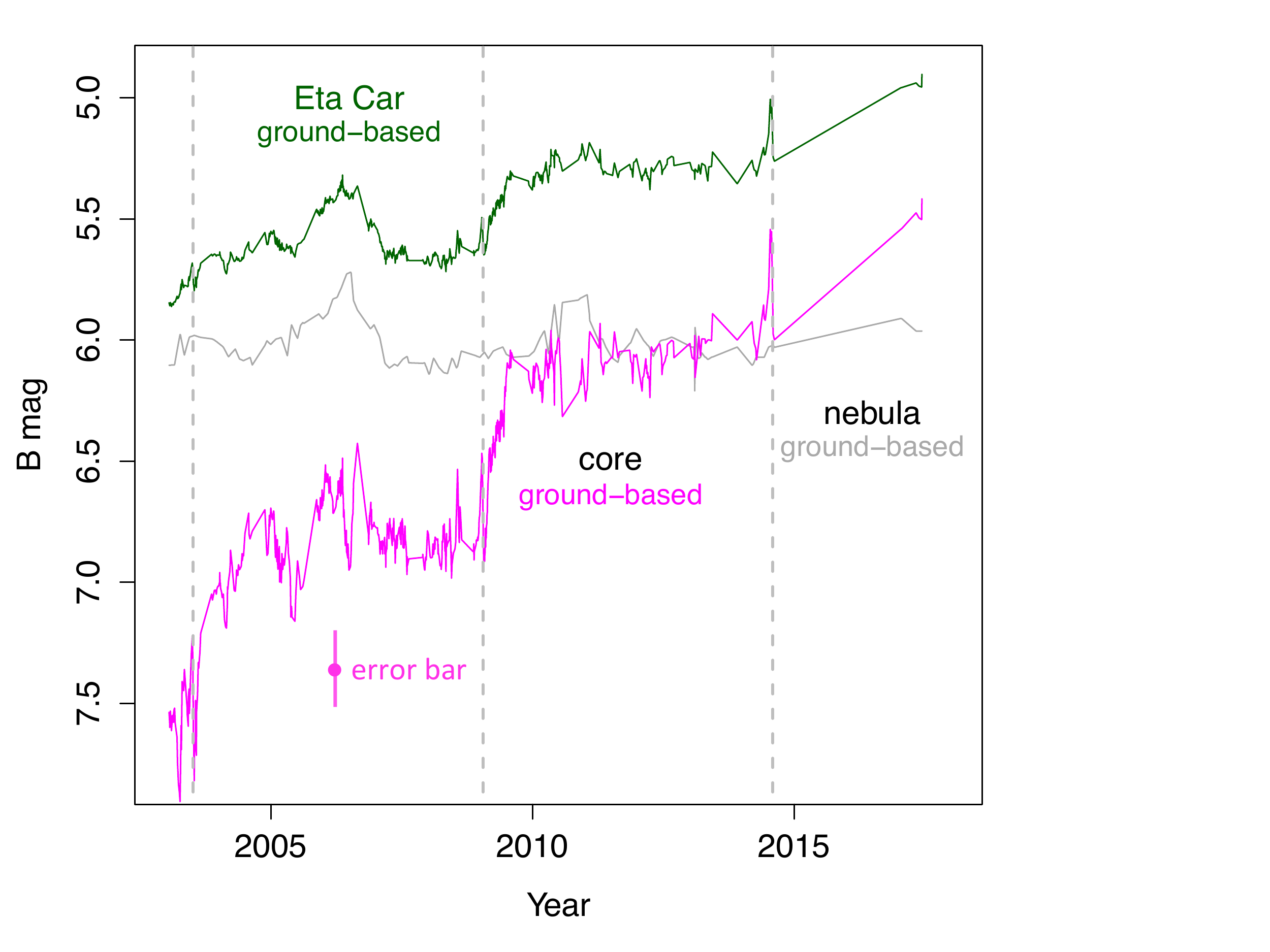}} \\[-3.0ex]
 \caption{$B$-band light curve of the core, the nebula and \ec. These have similar amplitudes as those in the $V$-band, although $\approx$\,0.5\,mag fainter. The calibration  is much worse than in the $V$-band because of using only a single epoch. The core's brightness dominated the nebula at some time after 2014.0. Typical error bar for the stellar core in the $B$-band is $\pm$\,0.1\,mag.}
 \label{B-lc} 
\end{figure}

The calibrated $B$-band light curves of the three apertures are similar to those in the $V$-band, but fainter by $\approx$\,0.6\,mag  (see Figure\,\ref{B-lc}). The uncertainty in the absolute calibration in $\rm B$ is larger than in $V$, but differential magnitudes are more accurate. 
The core became brighter than the nebula some time after 2014.0 in the $B$-band. This delay compared to that of the $V$-band is expected since the  (core)/(nebula) flux ratio decreases towards shorter wavelengths. 

\begin{figure}
 \centering
 \resizebox{\hsize}{!}{\includegraphics[width=\linewidth, viewport=0bp -15bp 600bp 520bp]{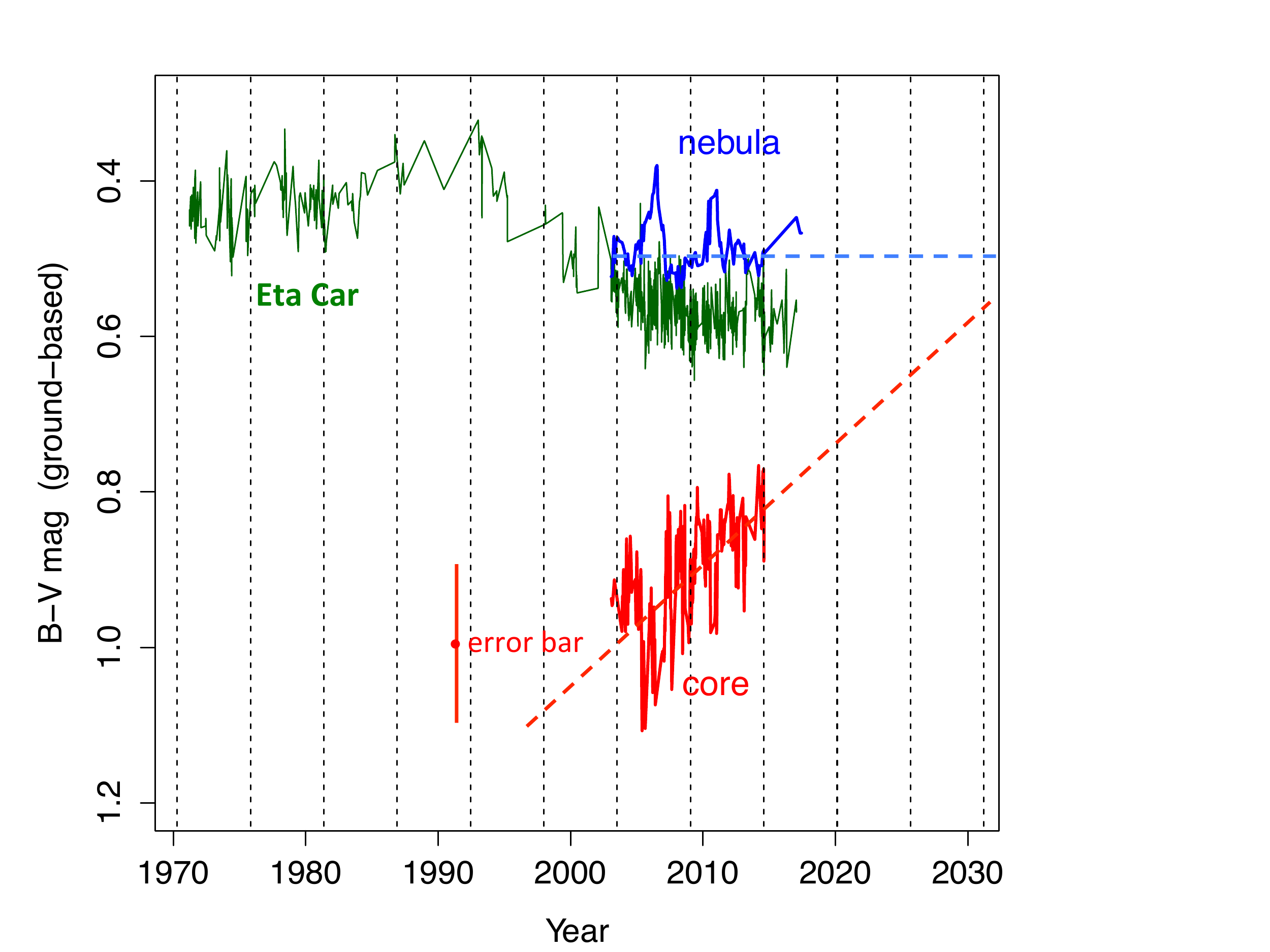}} \\[-3.0ex]
 \caption{Ground-based $B-V$ colour index of \ec\ since 1970; of the core (red) and the nebula (green) from 2003 to 2014.5. \ec\ blued slightly from 1970 to $\approx$1993 following the general evolution of the nebula. After that, it reddened at a faster pace, due to the fast brightening of the core, which is redder than the nebula.
 The evolution of the $B-V$ colour index of the core is $B-V$\,=\,0.85\,$-$\,0.013\,(\,${\rm year}$\,$-$\,2014.11\,), which will be equal to the nebula's colour index in 2042.  The typical error bar of the $B-V$ colour index of the stellar core is $\pm$\,0.15\,mag.
}
 \label{B-V_laplata} 
\end{figure}

Figure\,\ref{B-V_laplata} shows that the $B-V$ colour index derived from ground-based images has  a slow blueing increase of \ec\ from 1970 up to $\approx$1993. After that date, the slope in the $B-V$ curve became negative (redder) up to 2010. This occurred because the contrast between the core and the nebula increased and the core colour is redder than the nebula in this phase. Around 2010, the $B-V$ index of \ec\ flattened out because the core continued its blueing,
$\Delta(B-V)$\,=\,$-0.12$\,mag, in the interval 2003.5-2014.6). Since the core is getting bluer and the contrast with the nebula is rising, it is expected that the $B-V$ colour of \ec\ measured from the ground will soon  turn to  blueing.
 
There are small colour changes in the nebula brightness at mid-cycle, correlated with those in the core. They seem to be due to intrinsic colour variations in the core. The amplitudes of the $B-V$ colour changes in the core near periastron passage are not accurate, however, since the photometric model used to derive the nebula brightness assumes constant ratio between the outer and inner rings for all phases, in spite of small variability close to periastron passage.

	Figure\,\ref{B-V_laplata} shows the fit to the core and to the nebula $B-V$ colour index as a function of time. We binned the data a week at a time to reduce the scatter. The nebula is almost constant at $B-V$\,=\,0.48 in the time interval 2003-2014.5 and the core evolves in the form:
\begin{equation}
{B-V} = ( 0.85 \pm 0.01 ) - ( 0.013 \pm 0.001 ) (\rm{year} - 2014.11 ) 
\label{eqBV}
\end{equation}

\noindent
so that the core and the nebula are predicted to have equal $B-V$ in 2042. The uncertainty in the date is $\approx$\,10\,yr due to the large uncertainty in the $B-V$ colour index of the core ($\pm$0.1\,mag). However the changes in $B-V$ colour index depend upon complex physical effects that could change in the future. Nevertheless, the colour index of the core is blueing with time and that of the Homunculus nebula colour is constant, on average.  

\subsection{Ground-based $UBVIJHK$ light curves of {\ec}}
\label{sectionresults.5}

In this section, we work with the integral photometry of \ec\ because for many epochs and filters there is no way to separate the core from the nebula. Photometry of \ec\ is  challenging to interpret because it combines components that have different colours and fluxes. In previous sections, we showed that the core is brightening at a faster pace than the nebula. The colours vary by much less. We did not detect credible changes on an orbital time scale, although we showed that they occur on longer time scales at a lower level.

\begin{figure}
 \centering
 \resizebox{\hsize}{!}{\includegraphics[width=\linewidth, viewport=0bp 0bp 500bp 465bp]{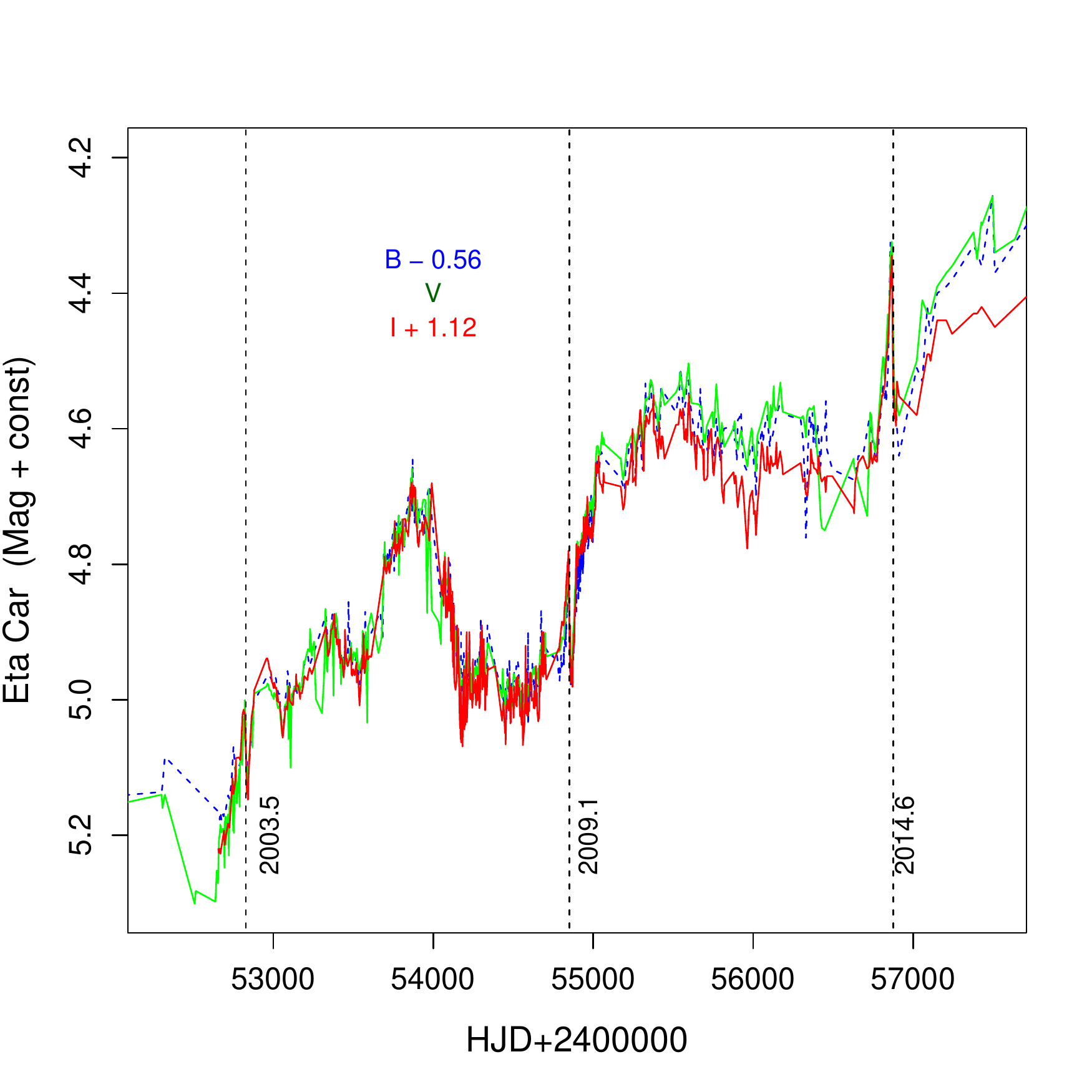}}\\[-2.0ex]
 \caption{Ground-based $BVI$ light curves of \ec. Colour index changes of \ec\ are very small in the time span of two cycles. Broad minima before periastron passage show up very clearly before periastron passages. They are colour-invariant and their depth seems to be variable from cycle to cycle.  The $B$- and $I$-band light curves were shifted vertically by adding a constant. Error bars are typically 0.005\,mag.}
 \label{BVI-lc} 
\end{figure}

The dense time-monitoring of \ec\ at La\,Plata enabled us to obtain a reasonably accurate comparison of light curves in different filters. As shown in Figure\,\ref{BVI-lc}, the colours are almost constant, with a small blueing over two cycles. This is not in contradiction with the observed blueing of the central source, since here we measured the integrated light of \ec, for which the flux from the core is diluted. The much lower blueing of \ec\ as compared to the central source indicates that the Homunculus lobes 
see different variations in the circumstellar dust than we see along our line of sight to the binary.
 
\begin{figure}
 \centering
 \includegraphics[width=\linewidth, viewport=60bp 30bp 620bp 520bp]{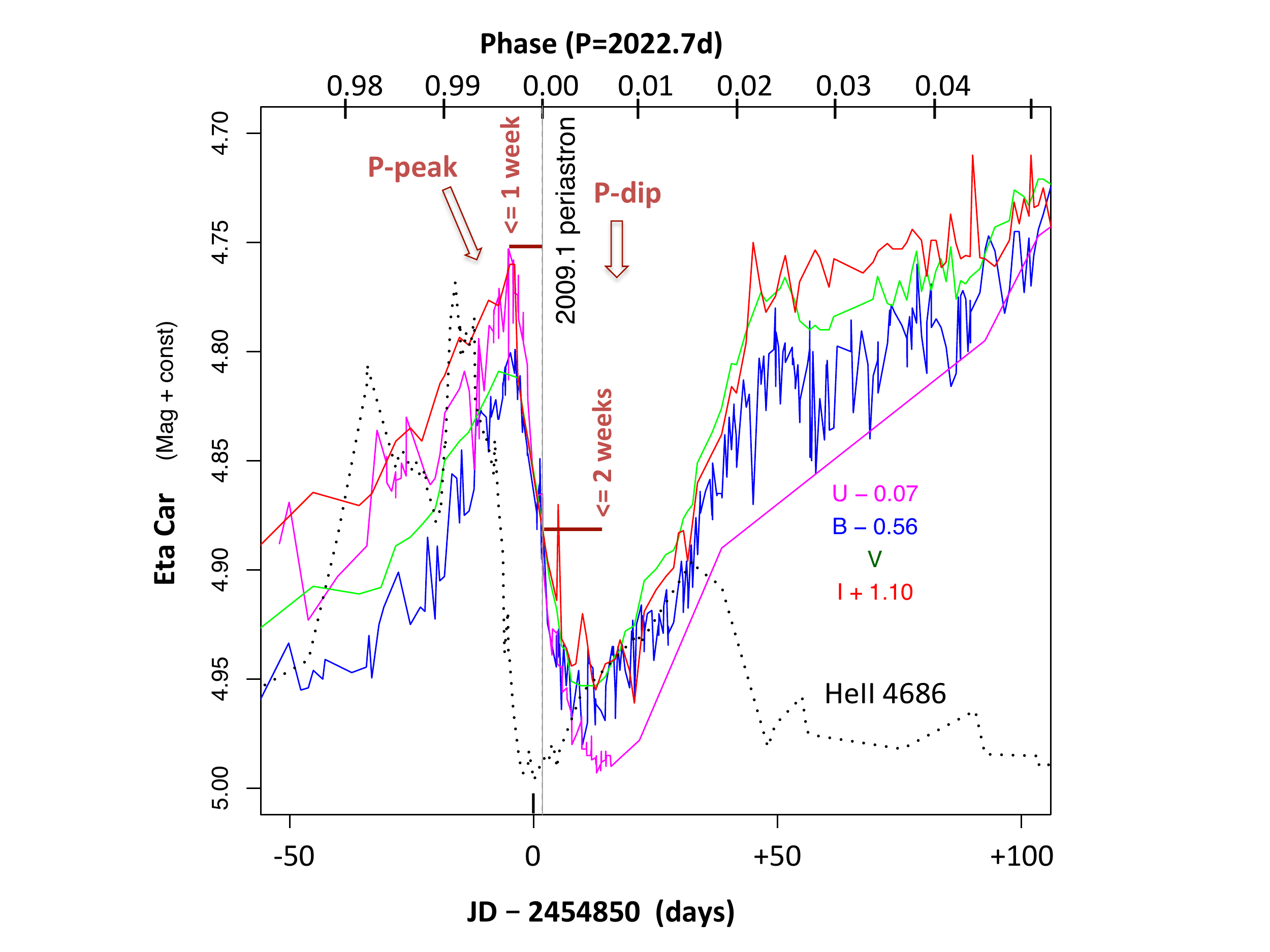}
 \caption{$UBVI$ light curves of \ec\ around the 2009.1 periastron passage (vertical line). The  P-peak started to rise some months before periastron passage and reached a maximum 5\,days before periastron passage. The P-dip drop in brightness lasted for 50\,days and was centred at 10\,days after periastron passage, 
 which is probably near the time of superior conjunction of the secondary. Magnitudes were shifted vertically to coincide at periastron passage. The intensity of the P-peak and depth of the P-dip are slightly dependent on the wavelength. The P-dip is a small feature when compared with the post-periastron minimum seen in $U$. The $\rm{He\,\textsc{ii}}$ 4686\,{\AA} line intensity is over-plotted, following \citet{Teodoro+2016}, but here the intensity is not to scale. Error bars are typically 0.005\,mag.}
 \label{peak-UBVI} 
\end{figure}

In Figure\,\ref{peak-UBVI} we present a zoomed view of the $UBVI$ light curves around the 2009.1 periastron passage. The minimum after periastron passage (P-dip) has a depth $\Delta$mag\,=\,0.17-0.25\,mag (wavelength dependent) and it lasts 50\,days, at orbital phases when the secondary star is ``behind'' the primary. It fits nicely in the ``bore hole'' effect scenario \citep{Madura12,Madura12a},  which enables our line-of-sight to view the interior of the wind-wind cavity for a brief time just before periastron passage -- see Figure\,\ref{Madura}. The peak before periastron passage (P-peak) has a small colour dependence in the $UBVI$ filters, which is due to the slightly different radii of the primary's wind photosphere. The P-dip arises as a result of a combination of a number of effects, the main one being the rapid wrapping of the wind-wind colliding region (WWCR) around the back side of the primary. The $\rm{He\,\textsc{ii}}$ 4686\,{\AA} line intensity curve is also shown (not to scale) for reference. A quantitative modelling of this effect requires the subtraction of the nebular contribution in different filters around periastron passage. Our data do not enable us to perform such a subtraction to the full extent, except for part of the $V$-band filter light curve, indicating that the P-dip minimum is $\Delta$mag\,$\approx$\,0.35 fainter than the P-peak.

The P-peak and the P-dip are  minor features in the photometric light curves, as can be seen in Figure\,\ref{BVI-lc}, and do not play any role in a global search for periodicity. However, when the data encompassing $\pm$\,2 months around periastron passage are selected and detrended for the long-term brightening, they are phase-locked and show a periodicity exactly equal to the spectroscopic period  \citep{Whitelock+1994, Lajus+2009}.

\begin{figure}
 \centering
 \resizebox{\hsize}{!}{\includegraphics[width=\linewidth, viewport=10bp 0bp 500bp 470bp]{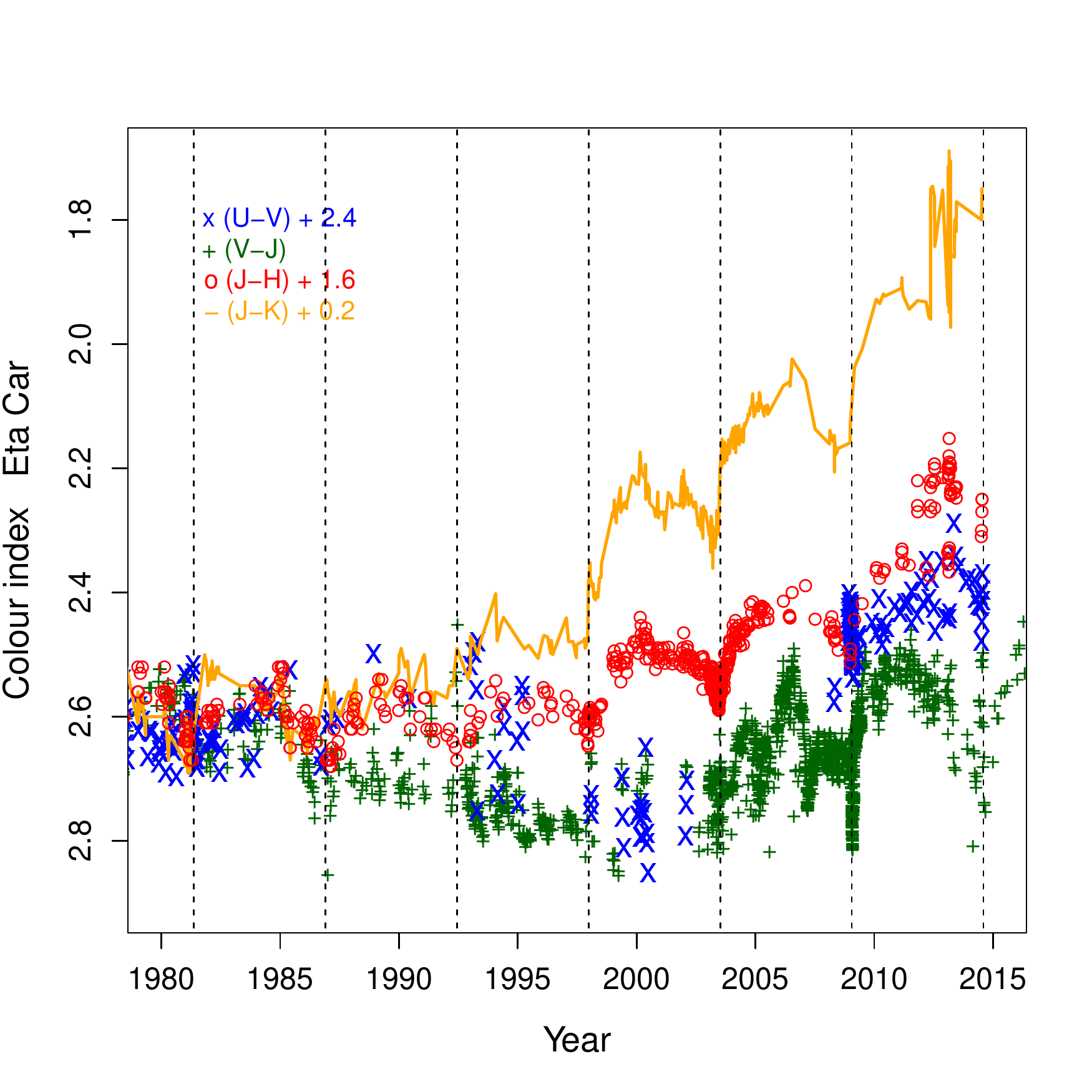}} \\ [-3.0ex]
 \caption{ $U-V$ (blue), $V-J$ (green), $J-H$ (red) and $J-K$ (orange) colour indices of \ec, shifted to overlap in 1980.0. Colours evolved to the red in the period 1980-2003 for colour indices involving optical bands, but remained constant in $J-H$ and blued in $J-K$. Differences are due to a  combination of fluxes from the nebula (bluer) and from the core (redder). Blueing (decreasing colour indices) affected all colours after 2003, when the contrast between the stellar and the nebular components was high. Error bars of the colour indices in the original measurements are typically smaller than $\pm$0.05\,mag, but after adjustments of zero points between different authors they can be twice as much on some occasions. A more realistic evaluation of the errors can be made by looking to the scatter of the points far from periastron passages.}
 \label{color-etacar} 
\end{figure}

Figure\,\ref{color-etacar} presents the evolution of colour indices during the last six cycles. All colours show a blueing after 2003.5, although the colour of the nebula remained almost constant.  Since the contrast between the star and the nebula grows towards longer wavelengths, the blueing is more significant for colour indices involving longer wavelengths. \citet{Mehner+2014} reported the same effect from NIR photometry and interpreted it as an increase in the temperature of the free-free emitting plasma close to the central binary. However, our results on a broad range of wavelengths indicate that a decrease in the foreground extinction towards the core plays the dominant role. Our interpretation is that the diminishing of extinction of the coronagraph exposes deeper regions of the primary photosphere due to penetration by the WWC cavity, where the temperature of the ionised gas is higher.  Colour indices changed  differently in the previous four cycles (1980-2003.5). The blueing rate was  minimal in $J-H$ and $J-K$, because the contrast between the core and the nebula was not sufficiently high. The $U-V$ and $V-J$ indices became redder with time during those cycles, which is easily explained by the fact that the core was redder than the nebula in the optical/UV and its brightness was increasing.
   
\section{Periodic and quasi-periodic photometric oscillations}
\label{sectionperiodicity}
\begin{figure}
 \centering
 \resizebox{\hsize}{!}{\includegraphics[width=\linewidth, viewport=70bp 40bp 620bp 520bp]{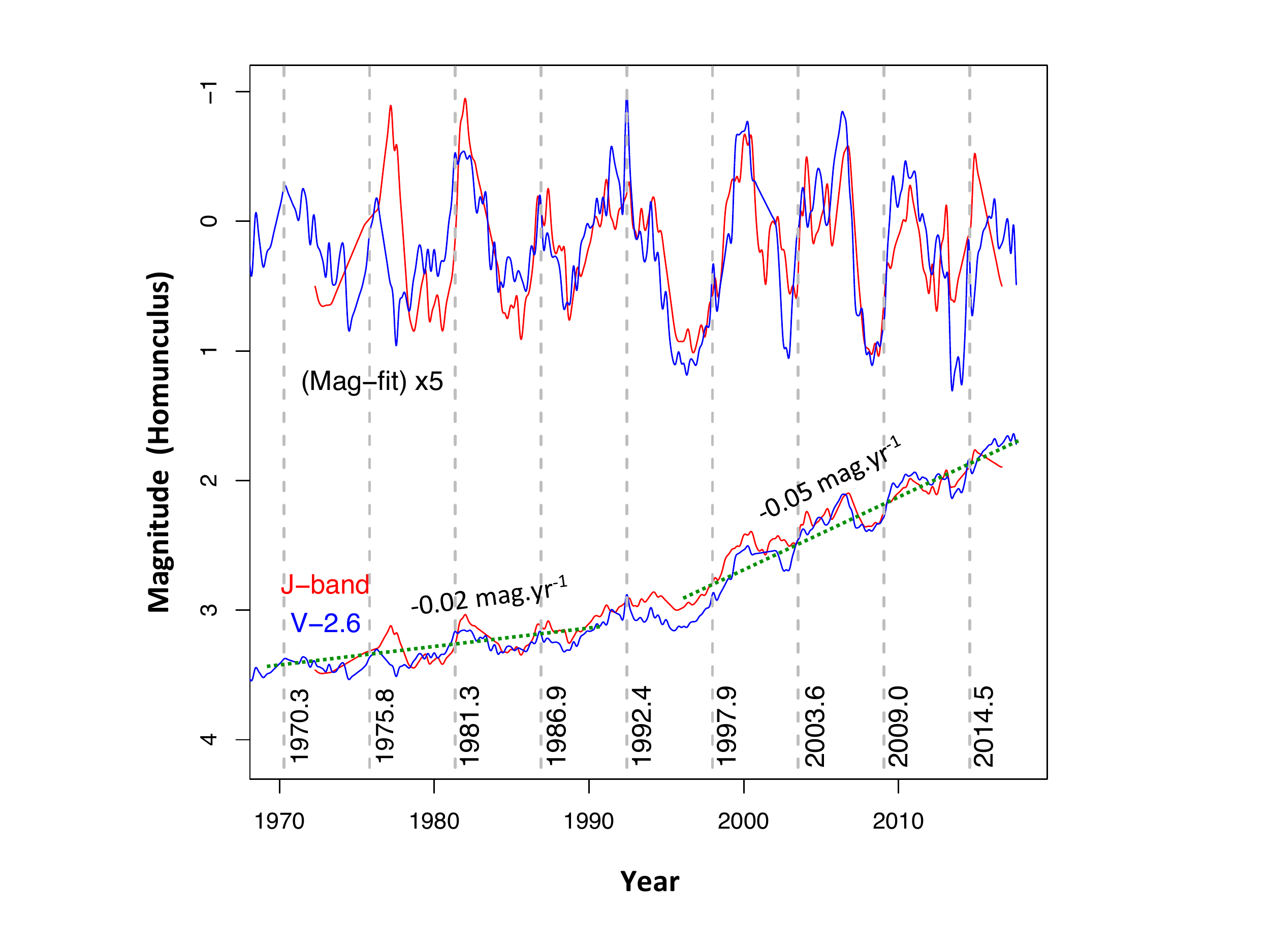}}
 \caption{{\it $a)$ Bottom}: {\ec} historical light curves in $J$- and $V$-band (offset by 2.6\,mag in $\rm V$) showing the two states of brightening, with a jump in the rate of brightening by a factor $>$\,2 in the early 1990's.
 $b)$ {\it Top}: $J$- and $V$-band light curves detrended from the long term brightening using  a low order curve in the period 1970-2018). Oscillations of $\pm$\,0.15\,mag are seen in both filters, almost in phase with the binary period. The minima are at their deepest before periastron passage, while they are already recovering at periastron passage. Typically the uncertainty is $\pm$\,0.03\,mag.}
 \label{VJhistoric}
\end{figure}

Strict periodicity compatible with the spectroscopic period ($P$\,=\,2022.7\,$\pm$\,0.3\,d) has been found in X-rays, NIR and optical light curves. We re-analysed the timing of the brightness peak near the periastron passage seen in Figure\,\ref{peak-UBVI} for the 2003.5, 2009.1 and 2014.6 periastron passages with the \textit{phase dispersion minimisation} method (PDM) and found P\,=\,2023.4\,$\pm$\,0.6\,d, in excellent agreement with the other methods. There are indications that other features in the light curve also have imprints of the orbital period. Figure\,\ref{VJhistoric} shows the light curves in $V$  and $J$ after correction for their long-term brightening trend obtained with the method detailed below. A periodic oscillation is clearly seen (multiplied by a factor of 5 in the plot), although it is not strictly phase-locked.

\begin{figure*}
 \centering
 \resizebox{0.7\hsize}{!}{\includegraphics[width=\linewidth]{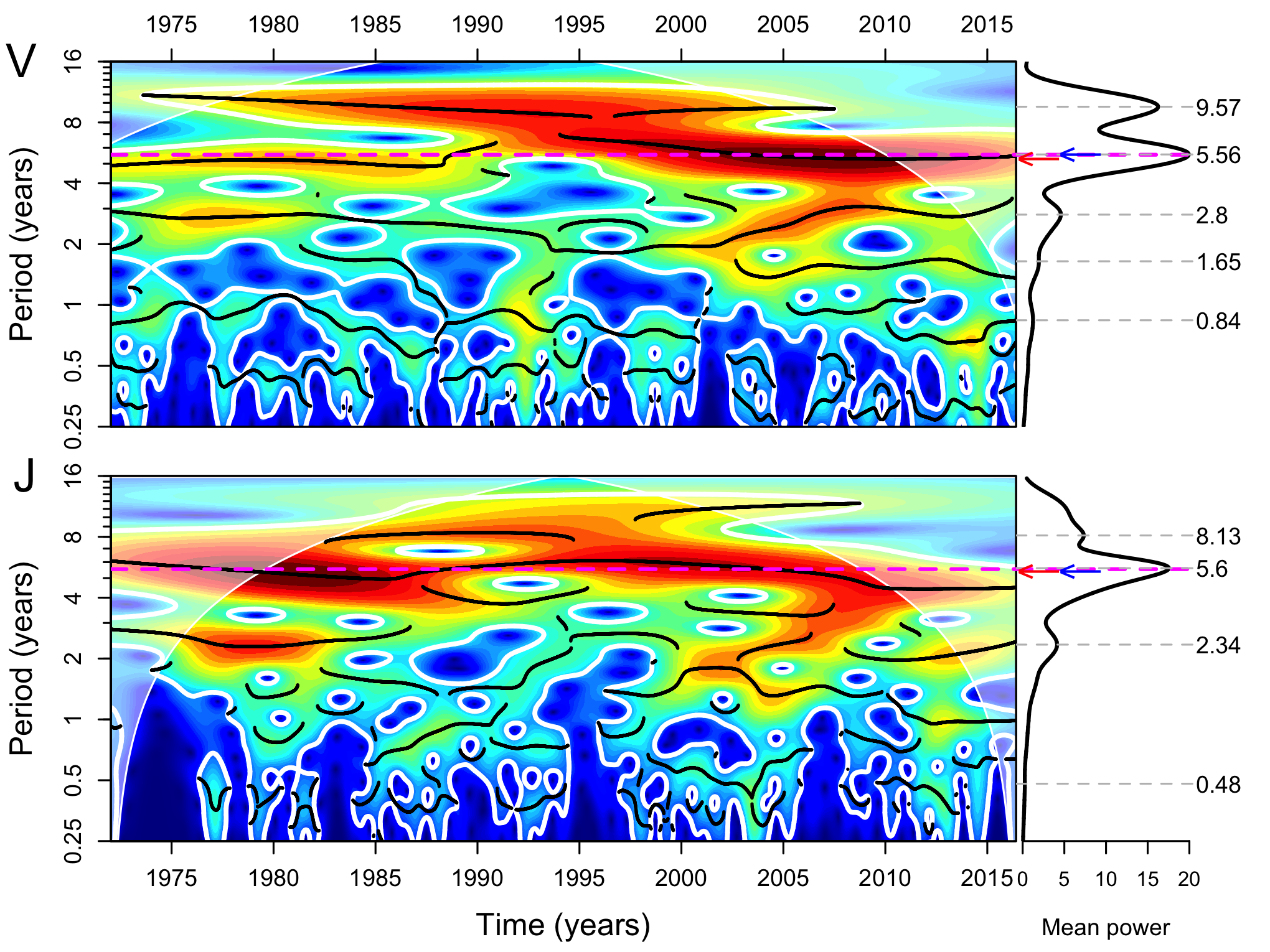}} \\ [-2.0ex]
 \caption{Wavelet spectrum of $V$- (top) and $J$-band (bottom) light curves of \ec.
 Significant periods across time are shown in the power spectrum (heat map) circled in white lines ($p$\,$<$\,10$^{-6}$), where the main instantaneous periods are represented by black lines corresponding to peaks in power (wavelet ridges). The mean power at the right was averaged across all time points outside the cone of influence (pale region where the estimation is prone to distortion), where the local peaks provide an estimate of the overall mean periods shown by dashed gray lines. The period with largest amplitude in both bands is the orbital modulation ($\sim$5.6\,yr) and varies in time similarly in $V$ and $J$, following closely the spectroscopic period (magenta dashed line -- $P$\,$=$\,5.538\,yr or 2022.7\,d). The main period detected with other methods closely matches the orbital modulation, shown as arrows at the right of the $y$-axis (Fourier in red and Lomb-Scargle in blue) for the $V$-band (Fourier\,=\,1979.7\,d and Lomb-Scargle\,=\,2083.8\,d) and $J$-band (Fourier\,=\,2026.5\,d and Lomb-Scargle\,=\,2026.4\,d). The average orbital modulation period is $P$\,=\,2030\,$\pm$\,13\,d. Discontinuities in the power spectrum and dark blue regions are often due to insufficient time sampling.
 }
 \label{wavelet-VJ}
\end{figure*}

The oscillations in $V$- and $J$-band light curves were analysed using the \texttt{CHUKNORRIS} pipeline \citep{DDamineli+2017}, which allowed us to estimate temporal changes in periodicity shown by the power spectrum in Figure\,\ref{wavelet-VJ}. First, the irregularly-sampled series were interpolated to a regular 1.5\,day sampling and then detrended using $loess$ (a local polynomial fit of second degree with a large moving window span\,$=$\,0.75) yielding a trend similar to a low degree polynomial. The detrended series was further filtered for periods longer than 96\,days and shorter than 6144\,days using a Discrete Wavelet Transform (Multi-resolution analysis with a Morlet wavelet), with periods over 6144\,days being considered as the remaining trend. Finally, the resulting filtered time-series (Figure\,\ref{VJhistoric}) was analysed with a Continuous Wavelet Transform (CWT) to estimate time-varying periodic components and compared with the overall periods detected by Fourier and Lomb-Scargle methods (Figure\,\ref{wavelet-VJ}).

	The Continuous Wavelet Transform \citep{TorrenceandCompo1998} detected regions in the time-frequency space with high power (shown in  red in Figure\,\ref{wavelet-VJ}), with significant periods estimated with an autoregressive process of order 1 as null model (implemented in the biwavelet package for R, \citealt{Gouhier2017}). The regions in the wavelet spectrum delimited by white lines are significant at $p$\,$<$\,$10^{-6}$ (Figure\,\ref{wavelet-VJ}), indicating a low probability that the detected periods are due to chance. The significant periods for each time point are shown by the wavelet ridges in black (Figure\,\ref{wavelet-VJ}), corresponding to the power peaks within the significant regions. The wavelet ridges indicate the existence of multiple periodic components consistently present through time. The center of the different components can be approximated by integrating the power spectrum in time and using the local peaks as reference, shown on the right $y$-axis (Figure\,\ref{wavelet-VJ}) considering only for the regions outside the cone of influence (where estimates suffer distortion).

    The main periods in $V$- and $J$-bands, $P_{\rm V}$\,=\,2029.44\,d and $P_{\rm J}$\,=\,2044.96\,d, respectively, are compatible with the spectroscopic period represented as a magenta dashed horizontal line (Figure\,\ref{wavelet-VJ}). The wavelet ridges in both bands indicated that the orbital modulation of the light curves was generally shorter than the spectroscopic period before 1989, in line with $P$\,$\approx$\,5.5\,yr reported by \citet{Whitelock+1994}. Between 1990-2005, the main ridges were longer than the spectroscopic period, becoming shorter than the spectroscopic period thereafter.
    The overall period was independently estimated with other methods that disregard changes in the orbital modulation, shown with the arrows along the right side of the $y$-axis (Figure\,\ref{wavelet-VJ}) for the Fourier transform (red) and Lomb-Scargle (blue) with P(Fourier)\,=\,1979.7\,d and P(Lomb-Scargle)\,=\,2083.8\,d for the $V$-band, and P(Fourier)\,$=$\,2026.5~d and P(Lomb-Scargle)\,$=$\,2026.4\,d for the  $J$-band. The average of these estimates (2029.7\,$\pm$\,13\,d) is consistent with the spectroscopic period (2022.7\,$\pm$\,0.3\,d). Furthermore, the cross-wavelet transform of the $V$- and $J$-bands matches the spectrum of the $J$-band alone, retaining all the main characteristics described above and supporting the similarity between both bands. The $V$-band series does not provide a unique period because the nebula is dominant and the light-travel time is significant to some parts of the reflecting bipolar caps. Although it would be possible to attempt more accurate period estimates from the broad-band optical data, it would not be of great use as the spectroscopic and X-ray periods are determined with far better temporal resolution.

	Figure\,\ref{starnebratio} shows that the core/nebula flux ratio increases to longer wavelengths so that in the $J$-band the core has been the dominant source of light over the nebula for many orbital cycles. The amplitude of the detrended $J$-band series has remained constant at $\Delta {J}$\,$\approx$\,0.3\,mag. The amplitude of the de-trended $V$-band series was $\Delta{V}$\,$\approx$\,0.1\,mag near 1970, and reached the same level as in the $J$-band, $\Delta{V}$\,$\approx$\,0.3\,mag, in 2010, when the core started to dominate the source of \ec's light in the $V$-band.
    There are two quasi-periodic components other than the main oscillation that appear consistently in both $V$- and $J$-bands, being the strongest at $\sim$\,3490\,d in the $V$-band and $\sim$\,2970\,d in the $J$-band (Figure\,\ref{wavelet-VJ}). However the $J$-band data are a shorter time series, which limits the detection of longer periods. The other period is centred at $\sim$\,1020\,d in the $V$-band and $\approx$\,860\,d in the $J$-band, but is significantly weaker. Since these periods could be suspected as simple harmonics of the orbital period, we used the discrete wavelet transform to isolate specific frequency bands and re-estimated the ridges confirming that they are independent periods. Moreover, if these other periods were simply harmonics, their ridges would have been practically parallel. Other weaker peaks in the mean power spectrum can be found at shorter periods, with corresponding ridges in the power spectrum; however, they are more variable and prone to artefacts derived from uneven sampling. The wavelet spectra have to be interpreted with care for short periods as the uneven sampling creates spurious discontinuities in the power spectrum, and consequently also in the ridges.

\begin{figure}
 \centering
 \resizebox{\hsize}{!}{\includegraphics[width=\linewidth, viewport=0bp 20bp 590bp 530bp]{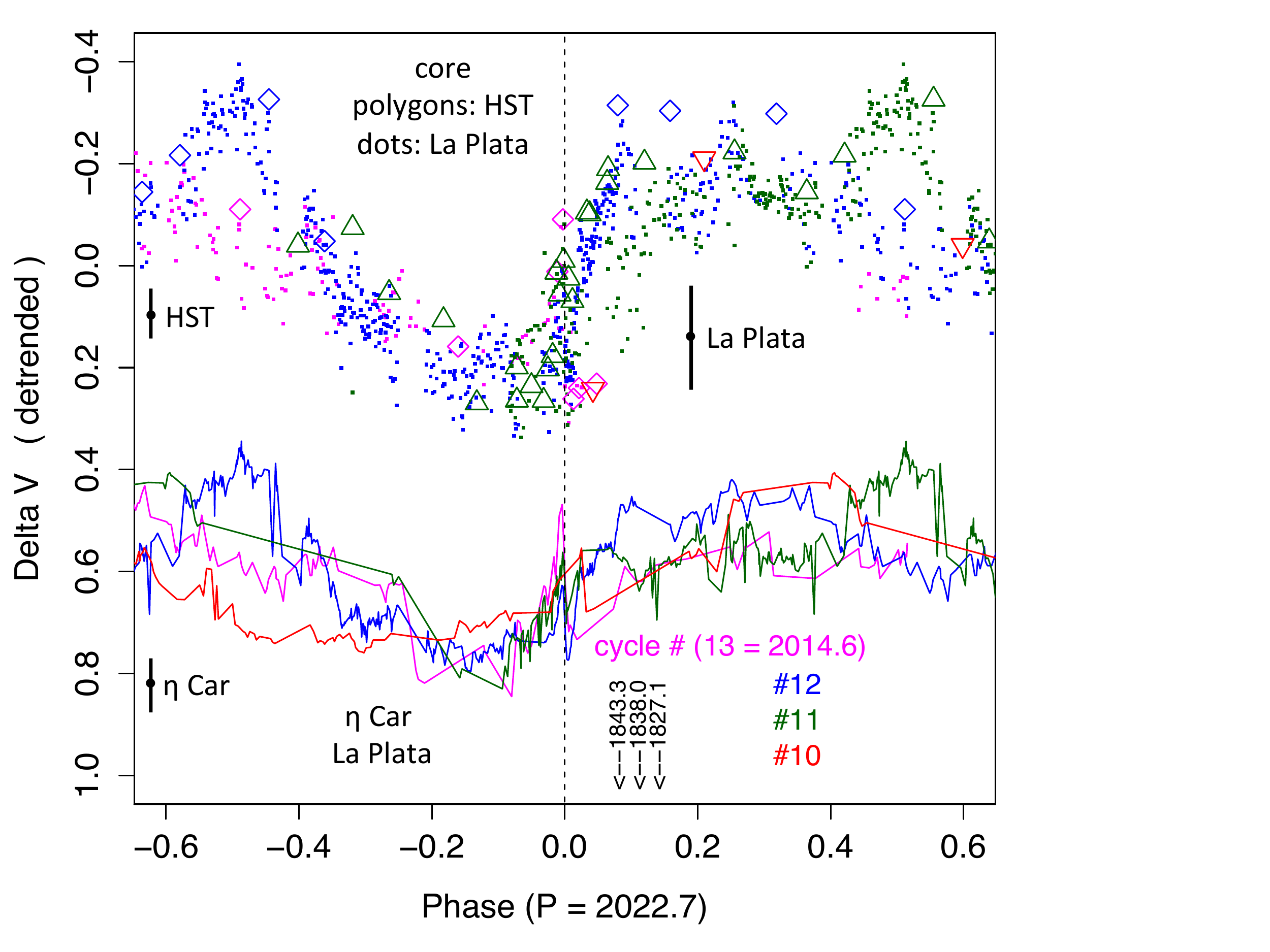}}
  \caption{Light curve of \ec\ and of the core folded with the orbital period ($P$\,=\,2022.7\,d) over the last four cycles, after detrending with a linear fit (as shown in Figure\,\ref{VJhistoric}). 
  $a)$ {\it Top}: the core component measured from HST images (polygons) and from the ground (points), showing an amplitude $\Delta{V}$\,=\,0.6\,mag, with error bar $\pm$\,0.1\,mag from the ground and $\pm$\,0.05\,mag from HST.
  $b)$ {\it Bottom}: the variation of the flux from \ec\ measured from the ground, with amplitude  $\Delta{V}$\,$\approx$\,0.3\,mag, with error bar $\pm$\,0.05\,mag. The variability of the de-trended light curve is significant at a level of $\sim$\,6\,$\sigma$ from La Plata and $\sim$\,12\,$\sigma$ from HST photometry. A vertical shift has been applied for visualization purposes. The three dates at the bottom mark the position of the pronounced peaks during the Great Eruption \citep{Smith+2011}. Cycle numbers follow \citet{Groh+2004}.}
 \label{MainOscillation}
\end{figure}

We used the orbital period to fold the $V$-band light curve (de-trended from the long-term brightening, but not filtered from other frequencies) for the last four cycles.   Figure\,\ref{MainOscillation} shows the phase-folded light curve measured from the ground and from the HST data for  \ec\ and the core (upper plot) measured from space (polygons) and from the ground (points). Cycle numbers follow the nomenclature by \citet{Groh+2004}; cycle 13 starts at the 2014.59 periastron passage. 
There are striking similarities between both sets of phase-diagrams, the main one being the existence of a deep and wide minimum, centred at $\approx$\,1\,yr ($\Delta\phi$\,$\approx$\,$-0.15$) almost a year before periastron passage. The amplitude of the core's light curve is $\Delta{V}$\,$\approx$\,0.6\,mag, as compared to $\Delta{V}$\,$\approx$\,0.3\,mag for \ec. The lower amplitude of the variation of \ec\ is due to the fact that the nebular contribution is almost constant. The deviations from phase-locked behaviour both in the light curve of the core and \ec\  are correlated and are more significant at mid-cycle. 

This large-amplitude oscillation can be assumed to be strictly periodic, with some disturbances by the nebula from a phase-locked behaviour (larger in $V$- than in $J$-band). 
The prime candidate for the intervening material producing a colour-independent variability is a dense plasma whose opacity is dominated by electron scattering 
associated with the colliding-wind cavity. 
 
\section{Extinction, rate of brightening and reddening law }
\label{sectionredlaw}

 The rate of brightening of the core can be derived from a linear fit of the magnitude decreasing in the time interval 2003-2015, which results in the equation: 
 \begin{equation}
V = 4.940 - 0.1134 ({\rm year}-2014.11)
\label{eqVpred}
\end{equation}
 
 By subtracting the un-reddened $V_{\rm 0}$\,=\,0.94\,mag  of the star, calculated by \citet{Hillier+2001} for the distance $d$\,=\,2.3\,kpc, the total $A_{\rm Vtot}$ extinction follows the equation:
\begin{equation}
A_{\rm Vtot} = 4.00 - 0.1134 ({\rm year}-2014.11)
\label{eqAvpred}
\end{equation}

We can isolate the extinction caused by the coronagraph by subtracting the other three sources of extinction, which we call ``foreground extinction'': $A_{\rm Vfg}$\,$=$\,$A_{\rm VISM}$\,+\,$A_{\rm Vcl}$\,+\,$A_{\rm VHom}$, in order that the total extinction is:
\begin{equation}
A_{\rm Vtot} = A_{\rm VISM} + A_{\rm Vcl} + A_{\rm VHom} + A_{\rm Vcoron}
\label{eqAvtot}
\end{equation}

\noindent where $VISM$ is the interstellar extinction; $Vcl$ is the intra-cluster extinction; $VHom$ is the contribution of the Homunculus wall in our sight line; and $Vcoron$ is the variable component we attribute to the dissipating coronagraph -- see subsection\,\ref{sectiondiscussion.3} below. 

\citet{Davidson+1995} reported the extinction towards the Weigelt clumps as $A_{\rm Vfg}$\,$\sim$\,1.5, which we adopt as a lower limit.  \citet{Hur+2012} using $UBVI$ photometry of the Trumpler\,16 and 14 (Tr16 and Tr14) stellar clusters did split the reddening in two components. The ISM component in front of the cluster with $R_{\rm VISM}$\,=\,3.1 has $E_{B-V}$\,=\,0.36 which results in $A_{\rm VISM}$\,$=$\,1.12\,mag. There is an intra-cluster component with a reddening law $R_{\rm Vcl}$\,=\,4.4\,$\pm$\,0.2 and $E_{\rm B-V}$\,=\,0.55 in the position of their map corresponding \ec. After subtracting the ISM component, the intra-cluster component of the colour excess is $E_{\rm B-V}$\,=\,0.19, which translates into $A_{\rm Vcl}$\,$=$\,0.84\,mag.
If we add $A_{\rm VHom}$\,$\sim$\,0.5\,mag to account for the Homunculus extinction, the foreground extinction would be  $A_{\rm Vfg}$\,$=$\,2.45\,mag. We adopt this value as an upper limit for the total ``foreground'' extinction, since no values larger than this have been reported.

We used 12 STIS spectra obtained in the time interval 1998.2-2015.7 selected by pairs in the same orbital phase, in order to check the consistency of our calculations.  We used the stellar continuum at the $\lambda$5495\,{\AA} narrow band, obtaining the brightening rate $\Delta5495$\,=\,$-$0.134\,$\pm$\,0.041\,mag\,yr$^{-1}$. Since the central wavelengths of the two filters ($V$ and $\lambda$5495\,{\AA}) are essentially identical, the slope of brightening is the same, so the transformation between $V$ and $\lambda$5495\,{\AA} magnitudes simply involves a constant.

Although this value is still compatible with the previous method that resulted in equation\,\ref{eqAvpred}, it has a larger error-bar. Using this narrow band and another one centred at $\lambda$4405\,{\AA}, we derive a reddening law of the form: $R_{5495}$\,=\,$A_{5495}$\,/\,$(A_{4405}-A_{5495})$. Since the orbital modulation is colour-invariant, the colour index $A_{4405}-A_{5495}$ does not suffer any change after subtracting the orbital modulation. We adapted the scheme of \citet{Fahed+2009}, using the ratio of variations in magnitude to those in colour. The derived average reddening law is $R_{5495}$\,=\,5.1\,$\pm$\,0.7 and since it is based just on the variable component, it represents the reddening law of the coronagraph. This reddening law implies large dust grains and is compatible with the low blueing gradient reported in Figure\,\ref{colormag-star}.

\section{Discussion}
\label{sectiondiscussion}

\subsection{General discussion}
\label{sectiongeneraldiscussion.1}

The method we designed for using the outer ring flux as a proxy for the contribution of the entire nebula successfully separated the core's flux from \ec\ in ground-based images. In the period 1998.2-2017.5, the core brightening was $\Delta$5495{\AA} or $\Delta{V}$\,$=$\,$-$0.113\,$\pm$\,0.002\,mag\,year$^{-1}$ and the same slope is valid for the $V$- and $B$-bands, while the nebular flux remained almost constant.
The core  blued at a very slow pace $\Delta$(4405\,{\AA}$-$5495{\AA})\,=\,$-$0.013\,$\pm$\,0.004\,mag\,year$^{-1}$, while the nebula colour index remained constant. The light curve of the core component shows that the stellar flux exceeded the nebular component in 2010 in the $V$-band and in 2014 in the $B$-band (see Figures\,\ref{V-all} and \ref{B-lc}). 
    
After being de-trended from the long-term brightening, the light curve shows a periodic oscillation close to the spectroscopic period ($P$\,=\,2022.7\,d) for all wavelengths. We call this the orbital modulation. The same effect as a function of time -- proximity to the spectroscopic period for more recent cycles -- is due to the increase of the contrast between the core and the nebula as a function of time. The amplitude of the orbital modulation is  $\Delta{V}$\,$\sim$\,0.6\,mag for the core and about half of this for the whole \ec, due to the dilution  by the nebular flux (see Figure \ref{MainOscillation}). The $V-J$ colour index of this oscillation seems to be constant (see Figure\,\ref{VJhistoric}). 
The significance of the light curve amplitude measured for the core is $\sim$\,6\,$\sigma$ from La Plata photometry and $\sim$\,12\,$\sigma$ from HST. 

We can rule out some mechanisms for the orbital modulation. Since the oscillation starts increasing in brightness approximately two months before periastron passage (when tidal effects are negligible), the maxima (which occur at mid-cycle) cannot be produced by periastron-passage induced mass ejections. Binary-induced pulsation of the primary star also can be ruled out because it would imply a huge radial pulsation: the stellar radius would change by $\approx$\,30\% from minimum to maximum, which would have a large impact on the X-ray light curve. A distortion in the shape of the primary star, caused by filling its Roche lobe, is also not a candidate mechanism since it would produce two maxima and two minima in a  binary period, which we do not see. 

\begin{figure}
 \centering
 \includegraphics[width=\linewidth, viewport=0bp -50bp 1050bp 480bp]{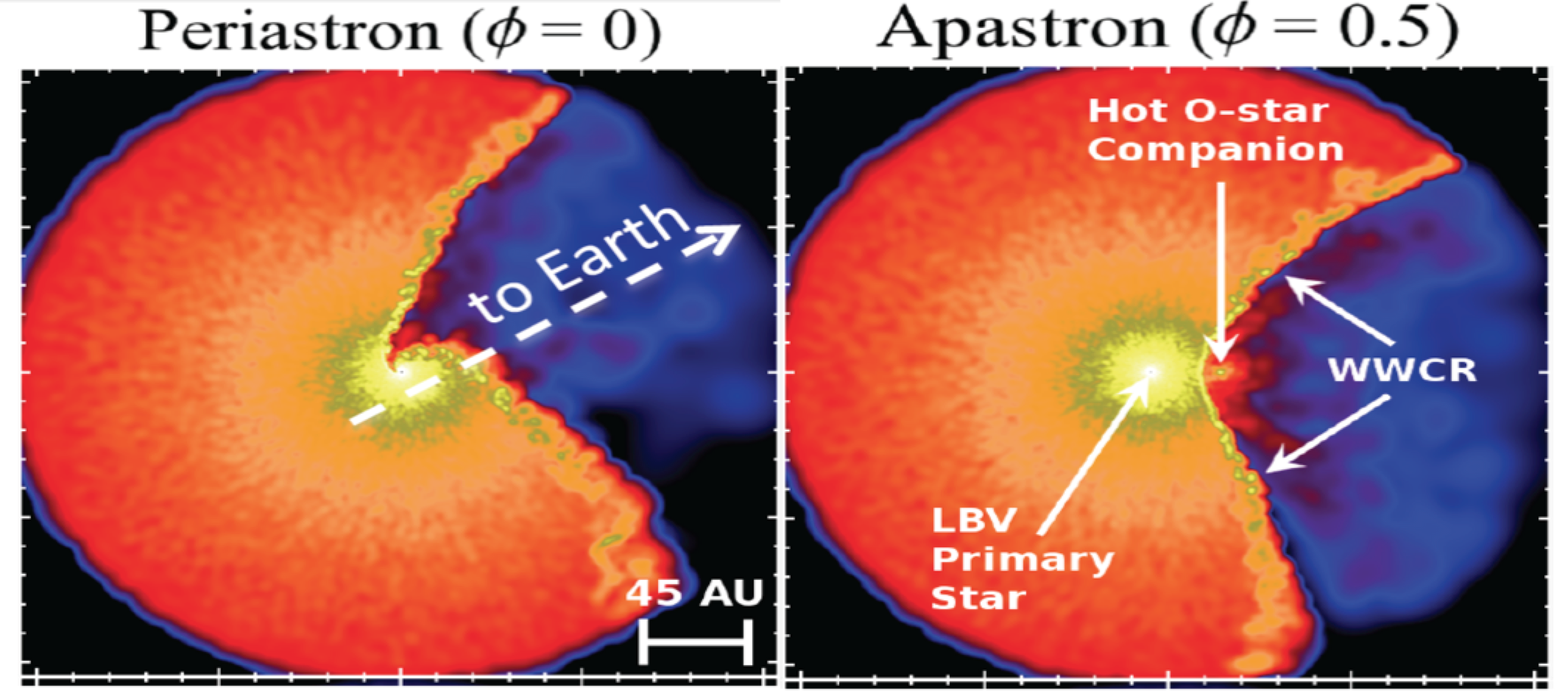} \\ [-3.0ex]
  \caption{Snapshots in the orbital ($xy$) plane at periastron and apastron phases ($\phi$, columns) from a 3D SPH simulation of the colliding stellar winds in {\ec} \citep{Madura+2013}. Colour shows density on a logarithmic scale. The locations of the stars and wind-wind collision region (WWCR) are indicated. The Earth's direction projected onto the orbital plane is shown in the left panel.}
 \label{Madura}
\end{figure}

 Our preferred explanation for Figure \ref{MainOscillation} is that the motion of the WWC-cavity as it rotates along the orbit projects different cross section areas in the direction of our line-of-sight -- see Figure\,\ref{Madura}. As a consequence,  combined with the fact that the WWC-cavity is not optically thin, this produces a light curve whose behaviour depends critically on the relative strengths of the stellar winds and on the orbital parameters. The large mass-loss rate of \ec\ and, as a consequence, the opacity in its wind makes this effect easier to identify in this object than in other colliding wind binaries. The best candidates to searching for this effect are LBVs in high state of S Doradus oscillations. A 3D SPH simulation to fine tune the relevant quantities in \ec\ will be done in a future work using the code described in \citet{Madura+2013}, \citet{Madura12} and \citet{Madura12a}.
 
It is plausible that the high-intensity peaks seen during the Great Eruption \citep{Damineli1996,Smith+2011} were enhanced versions of the P-peak displayed in Figure\,\ref{peak-UBVI}. Figure\,\ref{MainOscillation} shows the phases of the three bright peaks in the Great Eruption. They are shifted by $\Delta\phi$\,$\approx$\,+0.1 from periastron passage ($\approx$\,+200\,days), which might be caused by differences in the geometry of the shock cone at that time, since the momentum ratio between the two colliding winds was different then, because of the large density of the ejected matter.
The shift in phase between those three peaks indicates that the period was $\approx$3\% (2\,months) shorter than the present day orbital period \citet{Smith+2011}. Such a small change in the period would occur only if the mass lost by the primary star was replaced by a merger with a third star in the system, as discussed by \citet{Portegies+2016} and by \citet{Smith+2018b}. Since the mass lost in the great eruption was large ($>$\,12\,\ms\ following \citealt{Smith+2003} or $>$\,45\,\ms\ following \citealt{Morris+2017}), the putative merger star should have been very massive. Another possibility is that those peaks have no connection with the present-day P-peak and the closeness between them in orbital phase is merely a coincidence.

Some authors \citep{Mehner+2014, Davidson+1999} have claimed that periastron passage produces a ``jump'' in brightness, which would make the entire subsequent orbital cycle brighter. These apparent jumps at periastron passage, in reality, are produced by the long-term brightening plus the fast recovery branch of the P-dip across the periastron passage. The  detrended light curve  is sinusoidal-like, without periodic jumps. At periastron passage, the orbital modulation is already recovering from the minimum, which is centred at phase $\phi$\,$\approx$\,0.85 and maximum at mid-cycle.

At UV wavelengths, there is an additional minimum in the core's light-curve (the post-periastron minimum) which starts during the P-dip. Its depth and length increase rapidly towards shorter wavelengths. At 2200\,{\AA} its duration is t\,$\approx$\,2\,years and the depth is $\approx$\,1\,mag. This occurs when the WWC apex, which is a strong excitation source, is ``behind'' the primary star. The simplest explanation is that the P-dip is caused by the collective effect of the P-Cygni absorptions in permitted lines, as pointed out also by \citet{Martin+2006}. Such absorptions get deeper when periastron passage approaches and disappear slowly after periastron passage. They are pronounced in the UV, where $\rm{Fe\,\textsc{ii}}$ is dominant, and less important in the optical window. The recombination of gas that lies on the near side of the binary companion is probably located in the outer layers of the primary's wind. That region is partly hidden by the primary star itself and has lower excitation than the region irradiated by the light of the hot companion.

Once the $V$- and $B$-band magnitudes of the core recovered from the ground-based images were obtained independently, the local maxima appearing in the mid-cycle in the nebular light in Figures\,\ref{V-all} and \ref{B-lc} are likely real. The amplitudes are small ($<$\,0.2\,mag), but if they are produced by confined spots in the nebula \citep{Smith+2004a}, they should be obvious to detect in space-based images. We did not detect noticeable changes in the  outer/inner nebular ring ratio in ACS/HRC images, possibly because the time sampling was very sparse at mid-cycle. Figure\,\ref{B-V_laplata} shows that $B-V$ peaks ($\approx$\,0.1-0.2\,mag) in coincidence with the local maxima in the nebula. Although these peaks are redder than average, which suggests that they are reflecting intrinsic variations in the core, the core's $B-V$ light curve is too noisy to confirm this possibility.

We made a comprehensive study of the evolution in colour indices from $U$- to $K$-band covering more the 50\,years for \ec, as shown in  Figure\,\ref{color-etacar}. Regarding the NIR colours $J-H$ and $H-K$ our interpretation of the colour evolution disagrees with that suggested by \citet[][see their Figure\,2]{Mehner+2014}. Those authors claim that the observed blueing of the colour indices is due to an (intrinsic) increase in the temperature of the stellar wind which would indicate a decrease in the mass-loss rate. When comparing those colour indices with the optical colour indices, a different effect is clearly seen: the temporal evolution of the indices changes in a complex way as the contrast between the core and the nebula evolves. The contrast increases faster at longer wavelengths, making the colour indices involving NIR bands evolve much faster than those involving optical bands. The blueing started already in mid-1980 in the $J-K$ colour, two cycles later in $J-H$ and an additional two cycles later in $V-J$. The light seen at each wavelength comes from a different region of the system and is affected by different processes. For example, the $K$-band is affected by dust emission, in addition to ionised gas (free-free emission) and there might be different extinction for different wavelengths. In this way, the bluer colours at later times, shown in all colour indices, might be due to the exposure of deeper layers of the circumstellar gas \textcolor{magenta}{(where the gas is hotter)}, instead of \textcolor{magenta}{temporal} changes in the temperature at a given layer. Such a process is the same as that causing the long-term faint blueing observed in $U-V$ and $V-J$ (see Figure\,\ref{color-etacar}) during the last four cycles. Reddening of \ec\ in the $U-V$ and $V-J$ colour indices during the period 1980-2003 looks strange, but this can be easily explained. In the 1980's, the flux of the nebula was dominant over that of the core resulting in a corresponding blue colour for \ec. As the core  started to contribute more and more, the combined colour of \ec\ shifted to the red. However, the colour of the core is slowly evolving to the blue and, when it has  a significant impact at some specific wavelength, its colour evolution can dominate over that of the nebula.
  
 Our reddening law is in good agreement with one of the alternatives reported by \citet{Hillier+2001}. The value $R_{V}$\,$\sim$\,5.2 for the variable extinction indicates that the coronagraph has large dust grains. The absence of small size grains explains why the blueing is  weaker in the $UV$ (see Figure\,\ref{colormag-star}) as compared to the large brightening (see Figure\,\ref{stmag-star}) from cycle to cycle.

\subsection{The impact of the natural coronagraph on the core/nebula contrast and on the long term brightening}
\label{sectiondiscussion.2}

\ec\ can be compared to a Chinese lantern \citep{vanGenderen+1999}, in the sense that the Homunculus is a thin dusty surface that reflects the light of the central star. In fact, the spectrum we see looking directly towards the central source with HST is similar to that seen in reflected light: a B-type P-Cygni emission line star, with a rich set of H\,{\sc{i}}, He\,{\sc{i}} and Fe\,{\sc{ii}} emission lines. Because of strong extinction along our line of sight (i.e., the coronagraph) the spectrum we see from the ground is heavily contaminated by emission from the Weigelt clumps, the outer stellar wind  and the fossil winds \citep{HillierandAllen1992, Gull+2009, Davidson+1995, Hillier+2001b}. However, the contamination is decreasing as the coronagraph dissipates, and the relative intensity of the spectral lines as compared to the continuum flux is decreasing.These features, added to reports from previous works, as presented in the Introduction, are in line with the idea of a natural coronagraph in front of the central stellar object. The fast photometric brightening could happen because the coronagraph is dissipating or moving out of the line-of-sight. 

Although \ec\ is a dusty object, dust would evaporate quickly at a distance up to $\sim$200-500\,AU from the luminous primary star. In fact, \cite{Chesneau+2005} show that the immediate vicinity of the central object is empty of dust, forming the "Butterfly Nebula", whose borders are marked by dust emission at distance of a few hundred AU away from the central stars. The infrared imagery shown in Figure 1 of that work indicates considerable structure. The Weigelt clumps BCD  demonstrate that dust can exist at a distance of 0\farcs2-0\farcs3 (d\,$\sim$500-700\,AU)  from the central binary system \citep{Falcke+1996,Davidson+1997b}.

 In order to account for the fact that the BCD Weigelt clumps at 500-700\,AU are not covered by the coronagraph, it should have a diameter smaller than 500\,AU. These dusty BCD clumps have diameters $\sim$\,300\,AU, which brings the temptation to suggest that the coronagraph has the same nature.  Following \cite{Weigelt+1995} and \cite{Weigelt+2012}, the clumps were ejected in 1890 and are probably located near the Homunculus equator zone. But we do not know their exact location and the H\,$\alpha$ image reported by \cite{Weigelt+2012} plus the 190 and 307\,$\mu$m UV images reported by \cite{Weigelt+1995} show that additional, fainter clumps exist as different position angles and some of them might not be on the Homunculus equator. Then the coronagraph could be at $\sim$700\,AU or more in front of the binary system, if it is a clump similar as the BCD clumps.  Interestingly, the B clump, which is separated from the core by a projected distance of $\sim$\,300\,AU, was visible in the period 1985-2008, but dimmed substantially in recent years. This might be because it evaporated, which indicates dynamical changes in front of the central binary system.

Still another kind of object that could produce the same effect is the dusty 500x1500\,AU disk or bar  running NE-SW in front of the binary with width 500\,AU, found by \citet{Falcke+1996} based on speckle polarimetry. 

Unfortunately, the available observations do not yet allow us to decide which of the aforementioned possibilities  can best explain the nature of the puzzling coronagraph. However, our team hopes that our planned X-rays, HST/STIS and VLTI/MATISSE observations will improve our poor understanding.

We calculated the size of the occulting body required to explain the spectroscopic features (enhanced line wings) using the code described by \citet{Hillier+2001,Hillier+2006}. It must cover the formation region of the H\,$\delta$ line but leaving outside the formation region of H\,$\alpha$ and [Fe\,{\sc{ii}}] line wings. The maximum size is roughly the same as that of the Weigelt clumps, although it depends on the particular radial distribution of the extinction and even it does not need to be circular or centred on our line-of-sight. For a simplified exercise, we computed the spectrum arising inside and outside a radius r\,$=$\,0\farcs05 (r\,$\sim$110\,AU) from the primary star. We then divided the spectrum inside 0\farcs05  by 20 (dimmed by $\sim$3.3\,mag) and added the external part. 
[Fe\,{\sc{ii}}] lines are now clearly visible in the combined spectrum, and the EW of H\,$\alpha$ is enhanced by roughly 20\%. 

\subsection{The end of the brightening phase}
\label{sectiondiscussion.3}

\begin{figure}
 \centering
 \includegraphics[width=\linewidth, viewport=20bp 10bp 565bp 530bp]{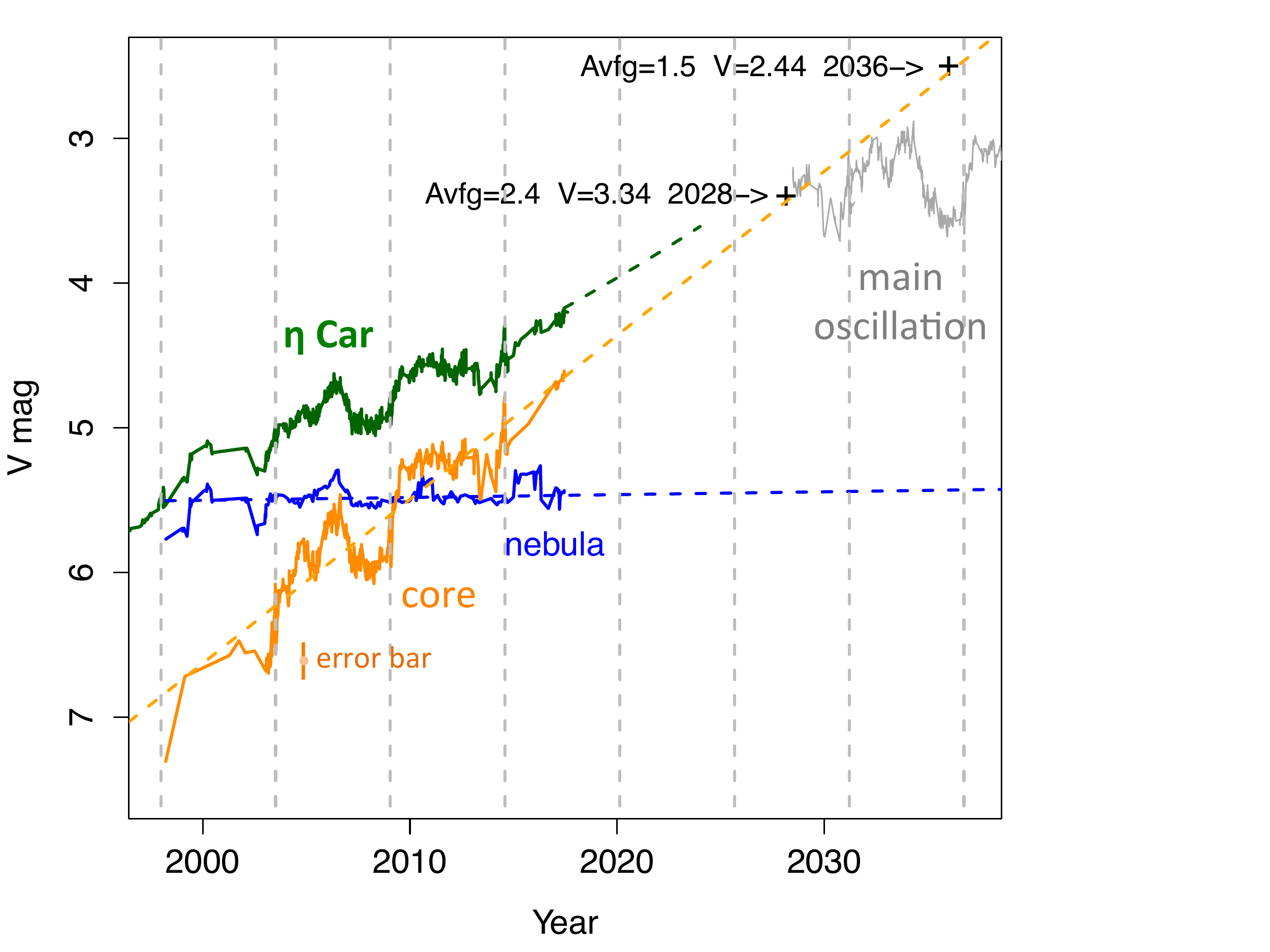}
 \caption{Prediction of the core (dashed orange) and of \ec\ light curve (dashed green) by extrapolating the linear trend fit to the core's brightness ($\Delta{V}$\,=\,$-$0.113\,mag\,yr$^{-1}$) and that of the nebula (blue lines: $\Delta{V}$\,=\,$-$0.002\,mag\,yr$^{-1}$). The green curve (\ec) is simply the sum of the blue and orange curves (fluxes converted to magnitudes). The black crosses along the predicted core's linear extrapolation (orange dashed line) mark the coronagraph's dissipation dates for the range of foreground extinctions ($A_{\rm Vfg}$\,=\,2.4\,$-$\,1.5 -- see text). The light curve in grey is a projection of the orbital modulation if the brightening stopped in 2028. Typical error bar is $\pm$\,0.1\,mag for the core.}
 \label{prediction} 
\end{figure}

The fact that the core of \ec\ is redder than the nebula and it is brightening faster causes the star to become brighter than the nebula earlier at longer wavelengths. As of early 2018, the core is already brighter than the nebular ejecta in the optical and NIR windows.  This results in ground-based spectra being similar to those which could be taken with STIS for those spectral windows in the near future. However, the nebula will continue to have an impact on \ec\ photometry for many orbital cycles. When the present phase of fast brightening ends ($>$\,2 cycles from now) the present brightening phase will resemble the 1941-45 jump in the optical light curve, although spread over a longer time-scale. The fact that the present-day dissipation of the coronagraph is also taking place on our side of the system suggests that the brightening and dissipation are similar or related events.  The important difference between the two events is that in the past case, the coronagraph continued producing enough extinction towards the star so that the high excitation lines appeared with large contrast. In the present situation, the eventual absence of the coronagraph will diminish the contrast between the nebular lines and stellar flux, causing a drop in the nebular-line equivalent widths (EWs).

The expected magnitude of \ec\ in the near future can be derived from Equation\,\eqref{eqVpred} or using the stellar un-reddened magnitude plus the total intervening extinction (Equation\,\eqref{eqAvtot}).   
Deriving the duration of the brightening phase is straightforward: It is obtained by (assuming that the central core remains constant and the brightening is due to dust dissipation) dividing the amount of extinction caused by the coronagraph  by the rate of brightening. The critical information is, therefore, the amount of extinction caused by the coronagraph and the rate of extinction fading, which is the inverse of the brightening rate. If the foreground extinction is underestimated, the correspondent for the coronagraph will be overestimated; the time spent in the brightening process to empty the reservoir of extinction will be longer; and the final brightness of the core will be higher when the light curve flattens out.

By using the foreground extinction in the  range reported in the literature: 1.5-2.4\,mag, the earliest date for the coronagraph extinction to disappear would be 2028 (when $V$\,=\,3.38) and the latest would be 2036 (when $V$\,=\,2.47) -- see Figure\,\ref{prediction}.
In  case the brightening ends earlier than 2028, the fainter brightness now could be explained if the foreground extinction is in reality larger than the upper limit $A_{V}$\,$=$\,2.4 we have adopted. In the case it ends later than 2036, the foreground extinction is smaller than $A_{V}$\,$=$\,1.5. The uncertainty in the predicted time is dominated by the error in the foreground extinction, not by the parameters of the fitting to the rate of brightening. In this way, the date of the end of brightening is 2032$\pm$4\,yr.

The magnitude at which the brightening phase will end brings an important piece of information about the process that occurred in the Great Eruption. Let us compare the luminosity of the system currently with that before the great Eruption. From the \citet{Hillier+2006} model, the primary star has $V_{\rm 0}$\,$\sim$\,1  and the secondary $V_{\rm 0}$\,$\sim$\,7\,mag. In  the case that the foreground extinction (ISM+intra-cluster) is $A_{V}$\,$=$\,2.4, the expected magnitude when the brightening phase ends will be $V$\,$\sim$\,3.4\,mag which is similar to that observed in the 1600's (within the uncertainty of $\pm$\,0.5\,mag). In the case that the foreground extinction at that time was the same as now, the present-day luminosity of the system is similar to that before the Eruption. However the magnitudes cannot be translated directly into luminosity, because the bolometric correction (and thus colour index) of the star in the 1600's is not known. In the case that the end date is after 2028, the final magnitude will be brighter than in the 1600's (at a rate of $\Delta{V}$\,$\sim$\,0.1\,mag\,yr$^{-1}$).
  
    The dissipation of the coronagraph has other important consequences. The contrast between the core and the nebular ejecta will be so large that many details, like the Weigelt objects, will fall below detectability. The line profiles and intensities of [Fe\,{\sc{ii}}] lines will change dramatically. The light curve of the core will continue to display the orbital modulation and lower amplitude quasi-periodic oscillations. The strictly periodic variabilities will continue to show up, maybe even becoming clearer than in the past: the $P$\,=\,58.8\,d ``pulsation'', the $P$\,=\,2022.7\,d orbital effects like the P-peak and its associated P-dip (in the optical window), plus the post-periastron minimum in the UV caused by recombination effects on our side of the primary soon after periastron. The system will continue to stay behind the interstellar plus intra-cluster extinction. The reflection-nebula ejecta are expected to continue brightening slowly and are not likely to become an H\,{\sc{ii}} region, at least in the near future, since the decrease of extinction affects mostly our line-of-sight. The high excitation lines will continue to be confined to the external opening of the wind-wind cavity, where we see the fossil-wind structures.

\section{Conclusions}
\label{sectionconclusions}

The analysis presented in this work gives observational support to the idea for a dusty clump (natural coronagraph) dissipating in our sight-line to the central star in \ec, which may be part of a larger obscuring structure.  The final V-magnitude of \ec\ after the dissipation of the coronagraph is important to know if the primary star was rejuvenated during the Great Eruption. If the brightening phase ends in $\sim$2036, the V-magnitude will probably be $\sim$\,1 magnitude brighter than in the 1600's. In the case the bolometric corrections (BC) were the same in the two phases, the central star would be 2.5 times more luminous now than before the Great Eruption. Even if there was a substantial mass ejection during the Great Eruption, the primary star could have been rejuvenated by internal mixing and/or by a stellar merger  \citep{Portegies+2016, Smith+2018b}.  However, the BC in the 1600's could have been, in principle, as extreme as the range 0 to -4.5 in order that \ec\ could be now $\sim$\,10 times more or up to $\sim$\,40 times less luminous than in the 1600's. Since the colour index of the central star was not observed at that time, the possible rejuvenation of the primary star cannot be well constrained.

The coronagraph is responsible for making the Homunculus an object of beauty, because it dims the central stellar core by many magnitudes enhancing the fascinating nebular features. In the case it disappears around 2036, for example, the core will be $\sim$\,10\,times brighter than the nebula, thus obfuscating the Homunculus  like in the case of nebulae around other LBVs.

The orbital modulation remained under-explored in the present work, since it demands  considerably more efforts which are being carried out through three-dimensional simulations of the wind-wind dynamics. Other colliding wind binaries should also show the same effect, depending on their orbital inclination, but such effect  has not been searched for in their historical light-curves. 3D simulations are expected to fit also the still mysterious P-peak, that we have attributed  in the present work to the ``bore hole'' effect \citep{Madura12a}. Both these features are conceivably produced by the wind-wind collision as it creates a spiral structure pattern orbiting around the centre of mass. Since the orbital modulation is colour-invariant, the WWC is not creating dust. As a matter of fact, the dust content of \ec\ is decreasing because of the drop in the infrared luminosity reported by \citet{Morris+2017} and by  \citet{Gaczkowski+2013}.
    
Nine components in the \ec\ light curve have been identified, six of them strictly periodic:
 \begin{itemize}
 \item{\it The long-term brightening:}
 \ec\ has been brightening at a pace of $\Delta{V}$\,=\,$-$0.02\,mag\,yr$^{-1}$ in the period 1945-90's which accelerated thereafter to $\Delta{V}$\,=\,$-$0.05\,mag\,yr$^{-1}$. This is due to the central star (core) which is contributing with $\Delta{V}$\,=\,$-$0.113\,mag\,yr$^{-1}$.  This corresponds to extinction decreasing at a rate of $\Delta{V}$\,=\,+0.63\,mag\,cycle$^{-1}$, while the star remains constant.
  \item{\it The orbital modulation:} 
A  colour-invariant periodic oscillation  ($P$=$P_{\rm orbital}$) with amplitude $\Delta{V}$\,=\,+0.6\,mag in the core component.  Its likely origin is the orbitally variable optical depth of the wind-wind collision cavity due to the bore hole effect. 
\item{\it The P-peak:}
A short peak in the light curve near periastron, produced by the bore hole effect \citep{Madura12,Madura12a}, when the apex of the WWC penetrates the He$^{+}$ layer of the primary star. This is connected to the orbital modulation.
\item{\it The P-dip:}
A minimum just following P-peak, with duration $\approx$\,2\,months and amplitude $\Delta{V}$\,=\,+0.35\,mag  for the core, present at optical and NIR wavelengths. It might be connected to the post-periastron minimum. The recovering branch of this minimum, added to the long-term brightening causes the false impression of a ``jump'' in the light curve after periastron passage. 
\item{\it The post-periastron minimum:}
A long (up to years) and deep minimum affecting UV light curves. This is produced by the opacity of Fe$^{++}$ recombining on our side of the primary star, when the ionising secondary star rotates to behind the primary. 
\item{\it The two short-period pulsations:}
Low-amplitude pulsations (milli-magnitude) with strict periodicity $P$\,=\,58.8\,d and $P$\,=\,22.7\,d, probably tidally excited \citep{Richardson+2018}. 
 \item{\it The two quasi-periodic oscillations:}
 variability in 2-3\,yr and 8-10\,yr time-scales with an amplitude of $\Delta{V}$\,=\,0.1-0.2\,mag. These might be due to intrinsic variability of the core, but difficult to be attributed to S\,Dor oscillations before confirmation from spectroscopy. 

\end{itemize}
Except for these two quasi-periodic oscillations, which are probably associated with the primary star, the other seven structures are reasonably well understood.

\ec's recent light curve is dominated by changes in its circumstellar ejecta. This is corroborated by the contrasting behaviour of the X-ray and radio light-curves, the former fairly repeatable and the latter variable from cycle to cycle. X-rays are emitted by the wind-wind collision and sensitive to the stellar parameters, while the radio emission is due to ionization of large volumes of gas, which is influenced by the way the UV radiation escapes through the circumstellar cavities, which are continuously changing. The star itself seems to have been reasonably stable. This is in contrast with the fame of \ec\ as an unstable star, which the object inherited from the Great Eruption.  Although that event happened relatively recently ($\sim$\,170\,years)  it seems to not have caused further instabilities in the primary star. On the one hand, the Kelvin-Helmholtz timescale of the primary of \ec~ is 100-1000\,years, depending on the assumption of its mass and radius, and such a scale is just a broad reference. In the observed examples, excess energy input to a star was thermalised sooner than predicted by theory. For example, V838\,Mon and V1309\,Sco suffered large bursts (they are thought to be stellar mergers, \citealt{Tylenda+2011}) involving much smaller radii and luminosities than in the putative case of an \ec\ merger. The Kelvin-Helmholtz time scale for these systems should be of the order of a million years, but the observed return to quiescence took just a few months. On the other hand, even if the light curve hints low amplitude variabilities of stellar origin, they seem to be unrelated to the Great Eruption since the release of the extra energy would cause a continuous decrease in the luminosity.

The smooth behaviour of the orbital modulation, obtained after detrending the long-term light curve does not support the idea of shell ejection at periastron. The frequently used eccentricity, $e$\,=\,0.9 \citep{Okazaki+2008,Teodoro+2016} seems to be close to its maximum value. Modelling of potential tidal interactions, beyond the scope of this paper, should follow up this conjecture.

 The lesson learned from this analysis is that, when interpreting the post-eruption phase light curve of unresolved LBV objects, the complex interplay of magnitudes and colours between the central star and its surrounding nebula must be taken into account. An important effect is that, like in \ec, the free-free emission from a dense wind makes the star red and the scattering process in a dusty nebula produces a bluer colour. If the light curve analysed here were coming from an object with an unresolved nebular contribution, it would probably be attributed to the star only, and would be impossible to model.

As a general conclusion, we did bring observational support to three characteristics of \ec\: a) the existence of a rapidly dissipating coronagraph; b) a large amplitude periodic photometric variability produced by the WWC-cavity; c) a photometric stability of the central star, higher than thought up to now. We did not modelled such features, which deserve to be explored in detail by subsequent theoretical and observational studies. 

\section*{Acknowledgements}
We thank Carlos E. Barbosa for making Figure\,\ref{figcolormap} and to N. Smith for comments on that Figure.
AD thanks to FAPESP for support (2011/51680-6).
LAA acknowledges FAPESP (2012/09716-6 and 2013/18245-0) and CAPES.
FN acknowledges FAPESP (2017/18191-8).
AFJM is grateful for financial aid from NSERC (Canada) and FQRNT (Quebec). NDR is grateful for post-doctoral support by the University of Toledo and by the Helen Luedke Brooks Endowed Professorship. TIM acknowledges partial support from HST program number HST-AR-14301, provided by NASA through a grant from the Space Telescope Science Institute, which is operated by the Association of Universities for Research in Astronomy, Incorporated, under NASA contract NAS5-26555. DSCD was funded by NSF, grant MCB 1616437/2016.

\label{lastpage}

\clearpage

\appendix
\section{HST photometry from ACS/HRC images and STIS spectra}
\label{appendix_a}

\begin{landscape}

\setlength{\tabcolsep}{9pt}
\begin{table}							
	\centering						
	\caption{Magnitudes of the core from ACS/HRC images and STIS spectra using synthetic photometry and narrow bands and of the nebula from ACS/HRC.}
    \label{tableHST-lc}						
	\begin{tabular}{ll|ccccccccl}
\hline
\multicolumn{1}{c}{year}	&	\multicolumn{1}{c}{phase}	&	\multicolumn{1}{c}{$\lambda$2255\AA}	&	\multicolumn{1}{c}{$\lambda$2520\AA}	&	\multicolumn{1}{c}{$\lambda$3363\AA}	&	\multicolumn{1}{c}{$\lambda$3660\AA}	&	\multicolumn{1}{c}{$\lambda$4405\AA}	&	\multicolumn{1}{c}{$\lambda$5495\AA}	&	\multicolumn{1}{c}{$\lambda$6800\AA}	&	\multicolumn{1}{c}{$\lambda$8000\AA}	&	\multicolumn{1}{c}{comm.}	\\
	&	\multicolumn{1}{c}{(*)}	&	\multicolumn{1}{c}{F220W}	&	\multicolumn{1}{c}{F250W}	&	\multicolumn{1}{c}{F330W}	&	\multicolumn{1}{c}{u-nb}	&	\multicolumn{1}{c}{b-nb}	&	\multicolumn{1}{c}{F550M}	&	\multicolumn{1}{c}{r-nb}	&	\multicolumn{1}{c}{i-nb}	& \\
\hline	
2002.7826	&	10.867	&	--			&	--			&	6.996\,$\pm$\,0.012	&	--			&	--			&	6.750\,$\pm$\,0.019	&	--			&	--			&	ACS*	\\
2002.7826	&	nebula	&	--			&	--		 	&	5.573\,$\pm$\,0.011	&	--			&	--			&	5.950\,$\pm$\,0.013	&	--			&	--			&	ACSneb	\\
2003.1151	&	10.928	&	7.850\,$\pm$\,0.031	&	7.080\,$\pm$\,0.012	&	6.973\,$\pm$\,0.014	&	--			&	--			&	6.773\,$\pm$\,0.019	&	--			&	--			&	ACS*	\\
2003.1151	&	nebula	&	--			&	5.525\,$\pm$\,0.009	&	5.535\,$\pm$\,0.009	&	--			&	--			&	5.717\,$\pm$\,0.012	&	--			&	--			&	ACSneb	\\
2003.1152	&	10.928	&	7.852\,$\pm$\,0.031	&	7.081\,$\pm$\,0.014	&	6.970\,$\pm$\,0.014	&	--			&	--			&	6.770\,$\pm$\,0.019	&	--			&	--			&	ACS*	\\
2003.1152	&	nebula	&	--			&	5.489\,$\pm$\,0.009	&	5.523\,$\pm$\,0.009	&	--			&	--			&	5.690\,$\pm$\,0.011	&	--			&	--			&	ACSneb	\\
2003.4451	&	10.988	&	7.965\,$\pm$\,0.032	&	7.050\,$\pm$\,0.014	&	6.754\,$\pm$\,0.012	&	--			&	--			&	6.539\,$\pm$\,0.017	&	--			&	--			&	ACS*	\\
2003.4451	&	nebula	&	--			&	5.457\,$\pm$\,0.009	&	5.739\,$\pm$\,0.011	&	--			&	--			&	5.555\,$\pm$\,0.011	&	--			&	--			&	ACSneb	\\
2003.4453	&	10.988	&	--			&	7.055\,$\pm$\,0.014	&	6.759\,$\pm$\,0.012	&	--			&	--			&	6.530\,$\pm$\,0.017	&	--			&	--			&	ACS*	\\
2003.4453	&	nebula	&	--			&	5.501\,$\pm$\,0.011	&	5.731\,$\pm$\,0.010	&	--			&	--			&	5.562\,$\pm$\,0.011	&	--			&	--			&	ACSneb	\\
2003.5466	&	11.006	&	8.536\,$\pm$\,0.042	&	7.469\,$\pm$\,0.017	&	6.869\,$\pm$\,0.013	&	--			&	--			&	6.536\,$\pm$\,0.017	&	--			&	--			&	ACS*	\\
2003.5466	&	nebula	&	--			&	5.827\,$\pm$\,0.010	&	5.522\,$\pm$\,0.011	&	--			&	--			&	5.618\,$\pm$\,0.011	&	--			&	--			&	ACSneb	\\
2003.5467	&	11.006	&	--			&	7.460\,$\pm$\,0.017	&	6.870\,$\pm$\,0.013	&	--			&	--			&	6.630\,$\pm$\,0.018	&	--			&	--			&	ACS*	\\
2003.5467	&	nebula	&	--			&	5.820\,$\pm$\,0.010	&	5.508\,$\pm$\,0.009	&	--			&	--			&	5.600\,$\pm$\,0.011	&	--			&	--			&	ACSneb	\\
2003.6994	&	11.033	&	--			&	7.552\,$\pm$\,0.018	&	6.862\,$\pm$\,0.013	&	--			&	--			&	6.384\,$\pm$\,0.016	&	--			&	--			&	ACS*	\\
2003.6994	&	nebula	&	--			&	5.698\,$\pm$\,0.009	&	5.361\,$\pm$\,0.011	&	--			&	--			&	5.614\,$\pm$\,0.011	&	--			&	--			&	ACSneb	\\
2003.6995	&	11.033	&	--			&	7.555\,$\pm$\,0.018	&	6.870\,$\pm$\,0.013	&	--			&	--			&	5.390\,$\pm$\,0.010	&	--			&	--			&	ACS*	\\
2003.6995	&	nebula	&	--			&	5.691\,$\pm$\,0.009	&	5.350\,$\pm$\,0.009	&	--			&	--			&	5.588\,$\pm$\,0.011	&	--			&	--			&	ACSneb	\\
2003.8687	&	11.064	&	8.374\,$\pm$\,0.039	&	7.236\,$\pm$\,0.015	&	6.597\,$\pm$\,0.012	&	--			&	--			&	6.301\,$\pm$\,0.015	&	--			&	--			&	ACS*	\\
2003.8687	&	nebula	&	--			&	5.667\,$\pm$\,0.009	&	5.470\,$\pm$\,0.011	&	--			&	--			&	5.635\,$\pm$\,0.011	&	--			&	--			&	ACSneb	\\
2003.8689	&	11.064	&	--			&	7.250\,$\pm$\,0.016	&	6.610\,$\pm$\,0.012	&	--			&	--			&	6.310\,$\pm$\,0.015	&	--			&	--			&	ACS*	\\
2003.8689	&	nebula	&	--			&	5.678\,$\pm$\,0.009	&	5.469\,$\pm$\,0.009	&	--			&	--			&	5.561\,$\pm$\,0.011	&	--			&	--			&	ACSneb	\\
2004.9300	&	11.256	&	7.420\,$\pm$\,0.025	&	6.456\,$\pm$\,0.011	&	6.258\,$\pm$\,0.010	&	--			&	--			&	6.085\,$\pm$\,0.014	&	--			&	--			&	ACS*	\\
2004.9300	&	nebula	&	--			&	5.517\,$\pm$\,0.009	&	5.528\,$\pm$\,0.011	&	--			&	--			&	5.710\,$\pm$\,0.011	&	--			&	--			&	ACSneb	\\
2005.5320	&	11.364	&	7.224\,$\pm$\,0.023	&	6.352\,$\pm$\,0.010	&	6.211\,$\pm$\,0.010	&	--			&	--			&	6.075\,$\pm$\,0.014	&	--			&	--			&	ACS*	\\
2005.8470	&	11.421	&	7.231\,$\pm$\,0.023	&	6.271\,$\pm$\,0.010	&	6.003\,$\pm$\,0.009	&	--			&	--			&	5.891\,$\pm$\,0.012	&	--			&	--			&	ACS*	\\
2005.8470	&	nebula	&	--			&	5.170\,$\pm$\,0.010	&	5.200\,$\pm$\,0.009	&	--			&	--			&	5.400\,$\pm$\,0.010	&	--			&	--			&	ACSneb	\\
2006.5880	&	11.555	&	7.166\,$\pm$\,0.022	&	6.149\,$\pm$\,0.009	&	5.850\,$\pm$\,0.008	&	--			&	--			&	5.757\,$\pm$\,0.012	&	--			&	--			&	ACS*	\\
2006.5881	&	nebula	&	--			&	5.439\,$\pm$\,0.009	&	5.380\,$\pm$\,0.011	&	--			&	--			&	5.550\,$\pm$\,0.011	&	--			&	--			&	ACSneb	\\
2007.0550	&	11.639	&	7.243\,$\pm$\,0.023	&	6.226\,$\pm$\,0.010	&	6.080\,$\pm$\,0.009	&	--			&	--			&	5.950\,$\pm$\,0.013	&	--			&	--			&	ACS*	\\
2007.0551	&	nebula	&	--			&	5.391\,$\pm$\,0.011	&	5.496\,$\pm$\,0.011	&	--			&	--			&	5.674\,$\pm$\,0.011	&	--			&	--			&	ACSneb	\\
1999.1400	&	10.210	&	8.426\,$\pm$\,0.134	&	7.387\,$\pm$\,0.083	&	7.119\,$\pm$\,0.073	&	7.150\,$\pm$\,0.074	&	7.341\,$\pm$\,0.081	&	6.896\,$\pm$\,0.066	&	6.522\,$\pm$\,0.056	&	6.247\,$\pm$\,0.049	&	Synt,2	\\
2001.2930	&	10.599	&	8.211\,$\pm$\,0.121	&	7.273\,$\pm$\,0.079	&	7.101\,$\pm$\,0.073	&	7.021\,$\pm$\,0.070	&	7.181\,$\pm$\,0.075	&	6.751\,$\pm$\,0.062	&	6.423\,$\pm$\,0.053	&	6.129\,$\pm$\,0.046	&	Synt	\\
2001.7480	&	10.681	&	8.020\,$\pm$\,0.111	&	7.050\,$\pm$\,0.071	&	--		 	&	--			&	7.180\,$\pm$\,0.075	&	6.650\,$\pm$\,0.059	&	--			&	--			&	Synt	\\
2002.0520	&	10.736	&	7.996\,$\pm$\,0.110	&	7.048\,$\pm$\,0.071	&	6.910\,$\pm$\,0.066	&	--			&	7.135\,$\pm$\,0.074	&	6.733\,$\pm$\,0.061	&	--			&	6.162\,$\pm$\,0.047	&	Synt	\\
2002.5050	&	10.818	&	7.885\,$\pm$\,0.104	&	6.969\,$\pm$\,0.068	&	6.908\,$\pm$\,0.066	&	6.879\,$\pm$\,0.066	&	7.175\,$\pm$\,0.075	&	6.720\,$\pm$\,0.061	&	6.361\,$\pm$\,0.052	&	6.118\,$\pm$\,0.046	&	Synt	\\
2003.1180	&	10.928	&	7.899\,$\pm$\,0.105	&	7.117\,$\pm$\,0.073	&	7.088\,$\pm$\,0.072	&	6.949\,$\pm$\,0.068	&	7.262\,$\pm$\,0.078	&	6.788\,$\pm$\,0.063	&	6.417\,$\pm$\,0.053	&	6.301\,$\pm$\,0.050	&	Synt	\\
2003.2380	&	10.950	&	--			&	--			&	--			&	--			&	7.223\,$\pm$\,0.077	&	6.742\,$\pm$\,0.062	&	--			&	--			&	Synt	\\
2003.3390	&	10.968	&	--			&	--			&	--			&	--			&	7.327\,$\pm$\,0.081	&	6.756\,$\pm$\,0.062	&	--			&	--			&	Synt	\\
2003.3750	&	10.975	&	8.025\,$\pm$\,	&	7.156\,$\pm$\,0.074	&	7.043\,$\pm$\,0.071	&	6.936\,$\pm$\,0.067	&	7.201\,$\pm$\,0.076	&	6.689\,$\pm$\,0.060	&	6.251\,$\pm$\,0.049	&	5.976\,$\pm$\,0.043	&	Synt	\\
2003.4150	&	10.982	&	7.953\,$\pm$\,	&	7.058\,$\pm$\,0.071	&	6.920\,$\pm$\,0.067	&	6.882\,$\pm$\,0.066	&	7.138\,$\pm$\,0.074	&	6.657\,$\pm$\,0.059	&	6.241\,$\pm$\,0.049	&	5.924\,$\pm$\,0.042	&	Synt	\\
\hline
	\end{tabular}
\end{table}
\clearpage

\begin{table}
\setcounter{table}{0}
	\centering
	\caption{\textit{(continued)} Magnitudes of the core from ACS/HRC images and STIS spectra using synthetic photometry and narrow bands and of the nebula from ACS/HRC.}
	\begin{tabular}{llccccccccl}
\hline
year&phase&$\lambda$2255\AA&$\lambda$2520\AA&$\lambda$3363\AA&$\lambda$3660\AA&$\lambda$4405\AA&$\lambda$5495\AA&$\lambda$6800\AA&$\lambda$8000\AA& \multicolumn{1}{c}{comm.}\\
&(*)&F220W&F250W&F330W&u-nb&b-nb&F550M&r-nb&i-nb& \\
\hline	
2003.4730	&	10.993	&	8.136\,$\pm$\,0.035	&	7.202\,$\pm$\,0.076	&	6.878\,$\pm$\,0.065	&	6.801\,$\pm$\,0.063	&	6.987\,$\pm$\,0.069	&	6.528\,$\pm$\,0.056	&	6.117\,$\pm$\,0.046	&	5.890\,$\pm$\,0.042	&	Synt,3	\\
2003.5070	&	10.999	&	--			&	--		 	&	6.913\,$\pm$\,0.067	&	7.143\,$\pm$\,0.074	&	6.990\,$\pm$\,0.069	&	6.457\,$\pm$\,0.054	&	6.117\,$\pm$\,0.046	&	--			&	Synt,4	\\
2003.5800	&	11.012	&	8.652\,$\pm$\,0.044	&	7.618\,$\pm$\,0.092	&	7.007\,$\pm$\,0.069	&	6.889\,$\pm$\,0.066	&	6.936\,$\pm$\,0.067	&	6.525\,$\pm$\,0.056	&	6.130\,$\pm$\,0.046	&	5.893\,$\pm$\,0.042	&	Synt,5	\\
2003.7230	&	11.038	&	8.599\,$\pm$\,0.043	&	7.552\,$\pm$\,0.089	&	6.862\,$\pm$\,0.065	&	6.672\,$\pm$\,0.060	&	6.726\,$\pm$\,0.061	&	6.334\,$\pm$\,0.051	&	5.958\,$\pm$\,0.043	&	5.700\,$\pm$\,0.038	&	Synt	\\
2003.8600	&	11.062	&	8.514\,$\pm$\,0.042	&	--		 	&	--			&	6.555\,$\pm$\,0.056	&	6.614\,$\pm$\,0.058	&	6.247\,$\pm$\,0.049	&	5.917\,$\pm$\,0.042	&	--			&	Synt,1	\\
2003.8780	&	11.066	&	--			&	--		 	&	--			&	--			&	6.603\,$\pm$\,0.058	&	6.213\,$\pm$\,0.048	&	--			&	--			&	Synt	\\
2004.1810	&	11.120	&	8.053\,$\pm$\,0.113	&	6.807\,$\pm$\,0.063	&	6.521\,$\pm$\,0.056	&	6.394\,$\pm$\,0.052	&	6.639\,$\pm$\,0.059	&	6.218\,$\pm$\,0.048	&	5.815\,$\pm$\,0.040	&	5.665\,$\pm$\,0.037	&	Synt	\\
1999.4450	&	10.265	&	--			&	--			&	--			&	--			&	--			&	6.881\,$\pm$\,0.066	&	--			&	--			&	nb	\\
2002.7800	&	10.867	&	7.973\,$\pm$\,0.033	&	--			&	6.996\,$\pm$\,0.069	&	--			&	--			&	6.850\,$\pm$\,0.065	&	--			&	--			&	nb,6	\\
2009.4950	&	12.080	&	--			&	--			&	--		 	&	--			&	--			&	5.326\,$\pm$\,0.032	&	--			&	--			&	nb	\\
2009.6310	&	12.105	&	--			&	5.986\,$\pm$\,0.043	&	5.535\,$\pm$\,0.035	&	--			&	--			&	--		 	&	--			&	--			&	nb	\\
2009.9310	&	12.159	&	--			&	--		 	&	--		 	&	--			&	--			&	5.273\,$\pm$\,0.031	&	--			&	--			&	nb	\\
2010.1700	&	12.202	&	--			&	5.802\,$\pm$\,0.040	&	5.447\,$\pm$\,0.034	&	--			&	--			&	--		 	&	--			&	--			&	nb	\\
2010.6340	&	12.286	&	6.270\,$\pm$\,0.030	&	5.700\,$\pm$\,0.038	&	5.250\,$\pm$\,0.031	&	--			&	--			&	--		 	&	--			&	--			&	nb	\\
2010.8170	&	12.319	&	--			&	--		 	&	--		 	&	--			&	--			&	5.149\,$\pm$\,0.030	&	--			&	--			&	nb	\\
2011.8860	&	12.512	&	--			&	--		 	&	--		 	&	--			&	--			&	5.180\,$\pm$\,0.030	&	--			&	--			&	nb	\\
2012.8000	&	12.677	&	--			&	--		 	&	--		 	&	--			&	--			&	5.200\,$\pm$\,0.030	&	--			&	--			&	nb	\\
2013.7030	&	12.840	&	6.050\,$\pm$\,0.020	&	5.560\,$\pm$\,0.036	&	5.300\,$\pm$\,0.032	&	5.100\,$\pm$\,0.029	&	5.390\,$\pm$\,0.033	&	5.183\,$\pm$\,0.030	&	5.108\,$\pm$\,0.029	&	--			&	nb,1,6	\\
2014.5300	&	12.989	&	5.748\,$\pm$\,0.019	&	5.260\,$\pm$\,0.031	&	4.957\,$\pm$\,0.027	&	4.733\,$\pm$\,0.024	&	5.104\,$\pm$\,0.029	&	4.913\,$\pm$\,0.026	&	--			&	--			&	nb,3	\\
2014.5800	&	12.998	&	6.411\,$\pm$\,0.021	&	5.830\,$\pm$\,0.040	&	4.969\,$\pm$\,0.027	&	4.734\,$\pm$\,0.024	&	4.967\,$\pm$\,0.027	&	4.807\,$\pm$\,0.025	&	--			&	--			&	nb,4	\\
2014.6200	&	13.005	&	7.212\,$\pm$\,0.024	&	6.760\,$\pm$\,0.062	&	5.502\,$\pm$\,0.035	&	5.159\,$\pm$\,0.030	&	5.301\,$\pm$\,0.032	&	5.169\,$\pm$\,0.030	&	--			&	--			&	nb	\\
2014.6640	&	13.013	&	7.100\,$\pm$\,0.024	&	6.580\,$\pm$\,0.057	&	5.398\,$\pm$\,0.033	&	5.123\,$\pm$\,0.029	&	5.305\,$\pm$\,0.032	&	5.145\,$\pm$\,0.029	&	--			&	--			&	nb,5	\\
2014.7100	&	13.022	&	7.005\,$\pm$\,0.023	&	6.280\,$\pm$\,0.050	&	5.324\,$\pm$\,0.032	&	5.055\,$\pm$\,0.028	&	5.267\,$\pm$\,0.031	&	5.117\,$\pm$\,0.029	&	--			&	--			&	nb	\\
2014.8570	&	13.048	&	6.031\,$\pm$\,0.020	&	5.720\,$\pm$\,0.038	&	5.162\,$\pm$\,0.030	&	4.911\,$\pm$\,0.026	&	5.239\,$\pm$\,0.031	&	5.087\,$\pm$\,0.029	&	--			&	--			&	nb	\\
2015.7050	&	13.201	&	5.600\,$\pm$\,0.019	&	5.390\,$\pm$\,0.033	&	4.960\,$\pm$\,0.027	&	--			&	5.094\,$\pm$\,0.029	&	4.974\,$\pm$\,0.027	&	4.750\,$\pm$\,0.025	&	--			&	nb,2 	\\

\hline
model&--&$-$0.329&0.279&$-$0.264&$-$0.099&0.399&0.938&1.502&1.797&--\\
band width&&50\AA&20\AA&50\AA&10\AA&5\AA&5\AA&50\AA&10\AA&--\\
\hline
	\end{tabular} \\
{Notes: Magnitudes are in the STMAG system; last column indicates the method of photometry (ACS\,=\,aperture photometry; Synt\,=\,broad band synthetic photometry on STIS spectra and nb\,=\,narrow-band photometry of STIS spectra.). Numbers in the last column indicate what pairs of measurements were used to derive the reddening law: $R_{5495}$\,=\,5.1\,$\pm$\,0.7. (*) Cycle numbering follows the scheme of \citet{Groh+2004} in which cycle 13 starts in 2014.59 and the period is P=2022.7d. Magnitudes from the model (row at the bottom) refer to the CMFGEN fit to the spectrum reported by \citet{Hillier+2001} for distance D\,=\,2.3\,kpc.}
\end{table}
\setlength{\tabcolsep}{6pt}

\end{landscape}

 \section{Study of the seeing impact on the nebular photometry }
 \label{appendix_b}
 
\begin{figure}
 \centering
 \includegraphics[width=\linewidth, viewport=0bp 25bp 490bp 525bp]{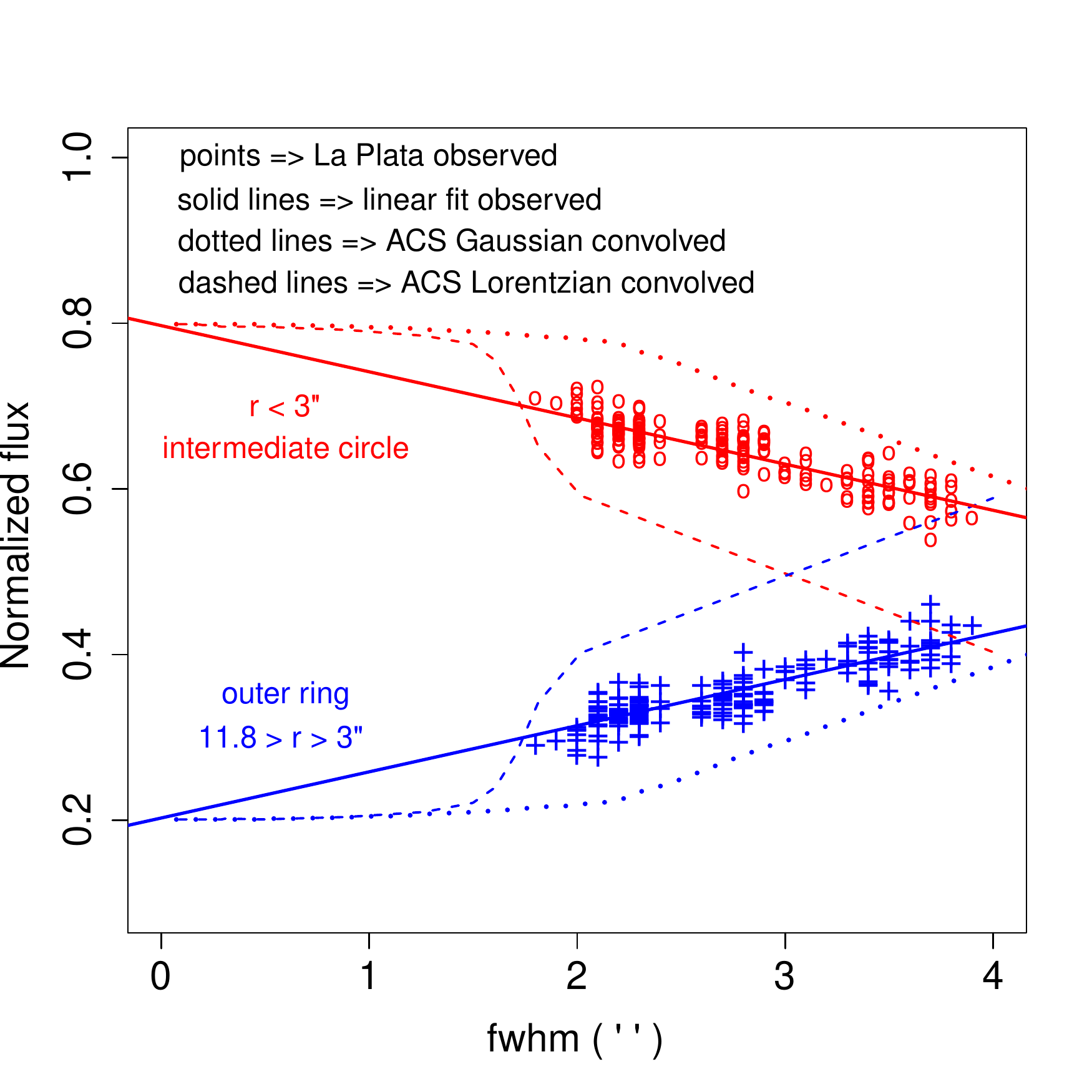} \caption{\textcolor{black}{Outer ring (blue) and intermediate circle (red) aperture fluxes measured in La\,Plata (points) $V$-band images taken on 20 January 2007 and in a coeval F550M ACS/HRC image convolved with Gaussian (dotted line) and Lorentzian profile (dashed line). Fits through fluxes of La\,Plata images are  represented by solid lines.  The plot indicates that the wings of the PSF in the ground-based images are more pronounced than in a Gaussian function and less than in a Lorentzian function. This is in agreement with \citet{Moffat1969} who reported  that the seeing profile is between these two extremes. The important result of these simulations is that in the ``bad seeing regime'' ($FWHM$\,$>$\,2\arcsec) the fluxes in the apertures we have used follow a linear relation with the seeing quality ($FWHM$ in this case).}}
 \label{seeingblur} 
\end{figure}

	In order to better understand the seeing effect in the intermediate circle($r$\,=\,$r2$) and in the outer ring   ($r2$\,$<$\,$r$\,$<$\,$r3$)  apertures, we convolved an ACS/HRC image taken on 20 January 2007 in the F550M filter with two different PSFs:
$a)$ a Gaussian; and $b)$ a Lorentzian, which has wings more enhanced than a Gaussian. A series of blurred images was produced and in each one the $FWHM$ was measured from a star in the field and the apertures were extracted in the aperture defined in Figure\,\ref{figcolormap}.
Fluxes were normalised to that of the outer circle aperture ($r3$\,=\,9\farcs5 for ACS/HRC images and $r3$\,=\,11\farcs8 for La Plata), which corresponds to \ec\ flux.
The same normalisation procedure was applied to  the ground-based images taken in the same date. 
In Figure\,\ref{seeingblur} the intermediate circle measurements are represented in red and the outer ring in blue. Fluxes from La\,Plata images are represented as circles and crosses. They are well fit by linear regressions (solid lines).

	In Figure\,\ref{seeingblur}, points are for a Gaussian blurring function on the ACS image and dashed lines for a Lorentzian profile. Although the Gaussian blurring reproduces the general linear relation between fluxes and $FWHM$ in the range of seeing sampled by the ground-based images, a function with stronger wings would be needed to reproduce the observed behaviour. In the case of a Lorentzian blurring, the fluxes also follow a linear trend for large $FWHM$. However, it removes too much flux from the intermediate circle to the outer ring, as compared to the observations. We do not discuss the regime of seeing $FWHM$\,$<$\,2\arcsec because it is dominated by the intrinsic PSFs and ground-based data in that regime are automatically rejected by our fitting procedure. We cannot do the exercise of convolution for $FWHM$\,$>$4\arcsec because the field star used to measure the $FWHM$ overlaps with a neighbour star.  The relevant result of this study is that the outer ring and intermediate circle fluxes of \ec follow a linear fit (as a function of the seeing parameter) in the ``bad seeing regime'' ($FWHM$\,$>$2\arcsec).
  
 \begin{figure}
 \centering
 \includegraphics[width=\linewidth, viewport=0bp 20bp 490bp 525bp]{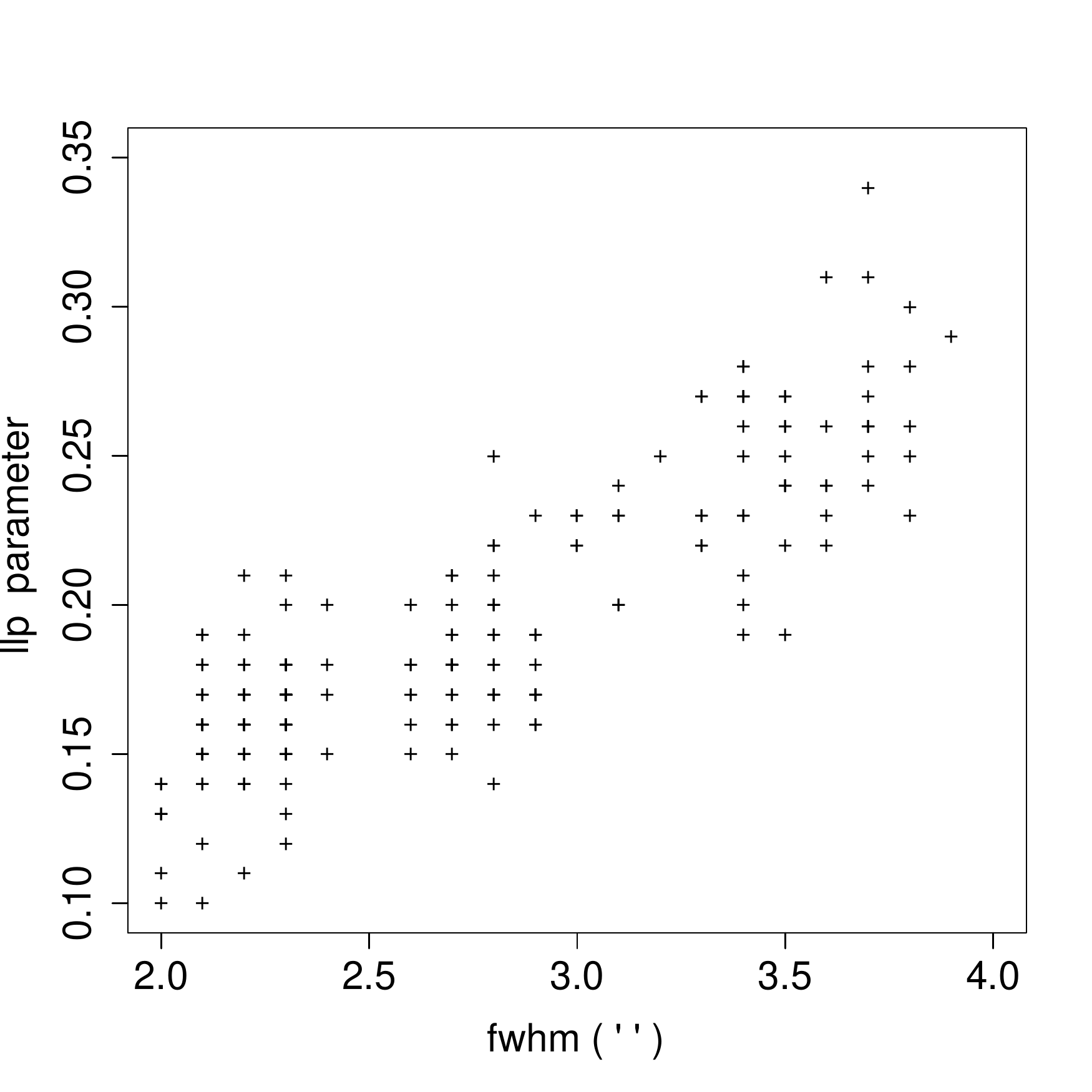} \caption{\textcolor{black}{Relation between the seeing parameter $FWHM$ and its proxy, the light loss parameter $LLP$ (defined by Equation\,\eqref{eqllf}) used to fit the fluxes inside the outer ring and inside the intermediate circle of \ec\ as a function of the seeing.}}
 \label{llpxfwhm} 
\end{figure}

  This study was not performed to fit the observations, for a number of reasons, in particular we do not know the PSF of the La\,Plata telescope and even if we did, it is not feasible to perform such a study for the many hundreds of nights. The goal is to check  if a linear fit of the flux as a function of the seeing parameter is reliable. Since this is the case, we could characterize every ground-based observing night by the intercept of the fits (at $FWHM$\,$=$\,0) and by the total flux of \ec, here used as a normalisation value. 
In Figure\,\ref{llpxfwhm} we show the relation between the measured seeing parameters: the $FWHM$ and its proxy, the light loss parameter ($LLP$) - see Equation\,\eqref{eqllf}. Measurements were made on the comparison star HDE\,303308 on 20 January 2007.

\end{document}